
\input phyzzx

%
%
%
%
%
%
\papers

\def\square{\kern1pt\vbox{\hrule height 1.2pt\hbox{\vrule width 1.2pt\hskip 3pt
   \vbox{\vskip 6pt}\hskip 3pt\vrule width 0.6pt}\hrule height 0.6pt}\kern1pt}

\def\BM{{\overline{\cal M}}}
\def\C{{\cal C}}
\def\N{{\cal N}}

\def\M{{\cal M}}

\def\O{{\cal O}}

\def\R{{\cal R}}
\def\T{{\cal T}}
\def\U{{\cal U}}
\def\V{{\cal V}}
\def\cut{\mathop{\rm cut}}
\def\supp{\mathop{\rm supp}}
\def\BR{{\bf R}}
\def\stub{{\cal S}}
\let\cd=\cdot
\let\e=\epsilon
\let\g=\gamma
\let\G=\Gamma

\let\lm=\limits
\let\ov=\overline
\let\p=\partial
\let\s=\sigma
\let\wh=\widehat
\let\x=\times
\overfullrule=0pt
\baselineskip 13pt
\pubnum{IASSNS-HEP 92/11 \cr
MIT-CTP-2072}
\date{February 1992}
\titlepage
\title{THE PLUMBING OF MINIMAL AREA SURFACES}
\author{Michael Wolf \foot{Partially supported by the
National Science Foundation;
Alfred P. Sloan Research Fellow.}}
\address{Department of Mathematics \break
Rice University \break
Houston, Texas\ 77251,\ U.S.A.}
\author{Barton Zwiebach \foot{Permanent address: Center for Theoretical
Physics, MIT, Cambridge, Mass. 02139. Supported in part
D.O.E. contract DE-AC02-76ER03069 and NSF grant PHY91-06210.}}
\address{School of Natural Sciences \break
Institute for Advanced Study \break
Olden Lane \break
Princeton, New Jersey\ 08540,\ U.S.A.}

\abstract{We study the metric of minimal area on a
punctured Riemann surface under the condition that all nontrivial
homotopy closed curves be longer than or equal to $2\pi$. By constructing
deformations of admissible metrics we establish necessary conditions
on minimal area metrics and a partial converse to Beurling's
criterion for extremal metrics.  We explicitly construct new minimal
area metrics that do not arise from quadratic differentials.
Under the physically motivated assumption of existence of the minimal
area metrics, we show there
exist neighborhoods of the punctures isometric to a flat semiinfinite
cylinder of circumference $2\pi$, allowing the definition of
canonical complex coordinates around the punctures.
The plumbing of surfaces with minimal area metrics
is shown to induce a metric of minimal area on the resulting surface.
This implies that minimal area string diagrams define a
consistent quantum closed string field theory.}
\endpage

\chapter{Introduction and Summary}

The motivation for the present work is a minimal area problem
for Riemann surfaces. Given a Riemann surface $\R$ the problem
asks for the conformal metric of least possible area under the
condition that all homotopically nontrivial closed curves on
the surface be longer than or equal to a fixed length, conventionally
taken to be $2\pi$ [Zw1]. The surface $\R$ is a
surface of genus $g \geq 0$ with $n \geq 0$ marked points,
or punctures, and the homotopy type of the curves is relative to
the punctures. The cases of Riemann spheres with one or no punctures
must be excluded since these surfaces have no nontrivial closed curves.

This generalized minimal area problem can be viewed as an
extremal length problem. As is well known, extremal length problems
in Riemann surfaces begin with the specification of a family $F$ of
curves in a surface [Ah,Beu,Ga]. The extremal length, which is a conformal
invariant, depends on the choice of the family $F$. The present
problem, as will be explained in \S2, is equivalent to an
extremal length problem where $F$ is the set of all nontrivial
closed curves in the surface. As such, the extremal length will be
a function of the moduli of the surface only.
This problem can also be viewed as a generalization of the minimal
area problems studied earlier by Jenkins and Strebel [Je,St]
where the length conditions apply to curves homotopic to a finite
set of curves, called an admissible set, containing non-intersecting,
non-homotopic, and nontrivial simple closed Jordan curves. Different
length conditions may apply for the different homotopy types.
In the present problem we impose the {\it same} length
condition on {\it all} nontrivial closed curves. \foot{It is
actually sufficient to consider all nontrivial {\it simple} Jordan
closed curves [Zw3], since nontrivial curves with
self intersections will satisfy the length condition
once all simple closed curves do.}  In other
words the unique length condition applies to curves homotopic
to a curve in the infinite set containing a representative from every
homotopy class. The curves in this set intersect and do not make
an admissible set.

The solution of the generalized minimal area problem is known for
all Riemann spheres ($g=0$; $n \geq 2$), namely, all surfaces
in the Riemann moduli spaces $\M_{g=0,n}$, and for a subset of every
$\M_{g,n}$ for $g\geq 1$ [Zw1]. The metrics arise from Jenkins-Strebel
(JS) quadratic differentials (quadratic differentials with closed
horizontal trajectories) with
second order poles at the punctures and with characteristic ring
domains that include a punctured disc around each puncture, and
a variable number of internal annuli. The horizontal trajectories,
all of which are of length $2\pi$, completely foliate the surface,
and are geodesics that saturate the length conditions.
We do not presently know the minimal area metrics for some subsets
of every $\M_{g,n}$ (with the exception of $g=1,n=0$). There was
some evidence that the minimal area metrics would not arise from
JS-quadratic differentials [Zw2], and in \S6 we describe such
an example.\foot{The specific metrics
conjectured in [Zw2] to give such an example are now known not to be
of minimal area.}

The relevance of the generalized minimal area problem for physics
arises in the context of developing a second quantized field theory
of closed strings [SaZw, KKS, KS]. The existence of such a field
theory demands, roughly speaking, that we find for every $\M_{g,n}$
(except $g=0;\, n=0,1$), a subset $\V_{g,n}$, and for each surface
in this subset we must specify a local coordinate $z_i$ around each
of the punctures (this coordinate is specified only up to a phase).
The subsets $\V$ define the vertices of the field theory. The
fundamental constraint they must satisfy is the following:
if we glue together the surfaces in the subsets $\V$, via the
plumbing relations $z_i w_i = t$ with $|t|\leq 1$, and obeying the
combinatorial rules of Feynman diagrams, we must generate
precisely the complete moduli spaces $\M_{g,n}$ [SoZw, Zw4]. The metrics
solving the generalized minimal area problem are expected to
determine the subsets $\V$ and tell us how to put coordinates
around the punctures of the corresponding surfaces [Zw2].
Basically, the minimal area metric is expected to be isometric
to a flat semiinfinite cylinder around each puncture,
this flatness requirement allowing us to define a canonical
coordinate. Since area is additive,
the plumbing of minimal area metrics is expected to induce in
the plumbed surface a minimal area metric. This is true if minimal
area metrics satisfy an amputation property [Zw2]:
amputation of the semiinfinite cylinder associated to a puncture
along a geodesic must induce on the truncated surface
a metric of minimal area. If plumbed surfaces acquire minimal area
metrics, the uniqueness of minimal area metrics [St,Zw1]
will imply that Feynman rules will not generate any surface
more than once. The $\V$'s are defined to be the surfaces that
are not generated by the sewing procedure. There is a more explicit
description of $\V$ in terms of the heights of foliations in
the metric of minimal area [Zw5].

The relevance of the generalized minimal area problem for
mathematics arises from the possibility of parametrizing
the moduli spaces of surfaces, and in particular, the
compactification divisors of moduli space in a canonical and geometrical
way. Now, the Deligne-Mumford compactification $\BM_g$ of
the moduli space of surfaces $\M_g$ ($g>1$) consists of a union
of the moduli space of surfaces with a compactifying set
of noded Riemann surfaces, i.e. connected complex spaces
where points have neighborhoods complex isomorphic to either
$\{ |z|< \epsilon \}$ (regular points) or $\{ zw=0;\,
|z|<\epsilon, |w| <\epsilon \}$ (nodes), and for which
each component of the complement of the nodes has negative
Euler characteristic (note that the complement of the nodes
is a collection of surfaces with paired punctures). The
topology of the compactification and a parametrization of
neighborhoods of points representing noded surfaces in
$\BM_g$ are given by the process of conformal plumbing or
``opening of a node'' (see [Mas],[EM],[Ber]); we describe
the process for a neighborhood in $\BM_g$ of a surface
$\R_0$ with a single node, leaving the general case to the
reader. To begin, for $|t|<\e $ we remove a small neighborhood
${\cal U} = \{ |z|<|t|, |w|< |t| \}$ of the node in the noded
Riemann surface $\R_0$ and form the identification
space $\R_t = (\R_0 - {\cal U})/<zw=t>$. We observe that
$\R_t$ is a smooth compact Riemann surface. The topology
of $\BM_g$ is then defined so that points in $\BM_g$ which
are near $\R_0$ are either those noded surfaces which
are quasiconformally close to $\R_0$, or those smooth surfaces
which are the result of opening the node (with small opening
parameter $t \not= 0$) of noded surfaces quasiconformally near
$\R_0$.

Now, these coordinates we have described are not canonically
defined, involving a choice of a neighborhood ${\cal U}$.
However, one of the goals in this paper is to show that, for
extremal metrics that are complete and sufficiently smooth in
a sufficiently large neighborhood of the puncture, we can make
the above process of opening the node somewhat more canonical:
there will be a maximal neighborhood of the puncture which is
flat and foliated by geodesics of length $2\pi$ homotopic to
the punctures. Taking this neighborhood to be a disk of radius
one, where the geodesics are the circles of constant radius,
we define $\R_t$ in a manner that depends only on the choice
of that particular point of the boundary of the disk that
should represent $z=1$. (In this choice, we must have an ambiguity
as to the $z$-argument of that point because of the topology of
$\T^*\BM_g$, see [W, Remark 4.1, p. 525]). Moreover, our
results imply that the minimal area metric on the smooth
surface $\R_t$, for $|t|<\exp (-\pi)$, is precisely
the metric induced on $\R_t$
by the identification process. Thus minimal area
{\it metrics} are consistent with the {\it conformal} process
of opening a node, and dually, with degeneration in
$\BM_g$ (pinching off a curve to obtain a noded surface).
In physics this translates into manifest off-shell
factorization of string amplitudes,
an essential property of covariant field theory.

Since minimal area metrics will not arise always from
quadratic differentials, we expect to find a natural and
interesting generalization of quadratic differentials.
We remark that there is presently no known
assignment of quadratic differentials
to all surfaces consistent with degeneration.
Moreover, the minimal area problem applies to surfaces
without punctures, where quadratic differentials do not yield
decompositions of the corresponding moduli spaces (see [Ha]).
The overall goal is for minimal area metrics to give a decomposition
of all moduli spaces. Each moduli space would be broken
into the pieces generated by the different patterns of foliations, these
pieces roughly correspond to the various Feynman graphs of the string
field theory. More explicit knowledge of the minimal area
metrics may be necessary to decide whether the decomposition is
actually a cell decomposition. This is the case for the
moduli spaces $\BM_{0,n}$ [Zw1].

The goal of the present work is to provide a mathematical
framework for the generalized minimal area problem.
We establish a useful partial converse to a criterion of
Beurling for extremal metrics. This amounts to necessary
conditions for extremal metrics.
In this paper we do not prove the existence of the minimal
area metric on an arbitrary Riemann surface. The above
results allow us to give, assuming the existence of a complete
minimal area metric smooth in some neighborhood of each puncture,
a proof of the requirements of flatness and amputation. This amounts
to establishing (modulo existence) that the minimal area metrics
define a quantum closed string field theory.

We have been able to obtain some minimal area metrics solving
our problem that do not arise from quadratic differentials.
The new property is that foliations by geodesics of
length $2\pi$ that cover the surface must intersect over regions
of non-zero measure. Since these metrics are quite
novel, and provide further evidence of existence of the minimal
area metric for arbitrary Riemann surfaces, we will
present them in \S6.

The contents of this paper are organized as follows. In \S2 we
discuss preliminaries and notation. We explain the relation with
extremal length and the notion of saturating geodesics. It is
shown that for a complete minimal area metric, smooth and nonvanishing
in some neighborhoods of the puctures, we have a foliation by
saturating geodesics around the punctures. In \S3
we prove a lemma that amounts to a (partial) local converse to
Beurling's criterion [Beu]. While Beurling's criterion is a sufficient
condition for a metric to be of minimal area, our converse gives
necessary conditions. Roughly speaking, the necessary conditions
apply if a minimal area metric
$\rho_0$ is smooth on a compact domain which is foliated by
saturating geodesics $\g_i$. Then, for any smooth $h$ in this domain
such that $\int_{\g_i} h |dz| \geq 0$, one
must have $\int h\rho_0 \hbox{dxdy} \geq 0$. Here the saturating
geodesics may belong to a finite number of homotopy classes of
curves. Using this criterion
we show flatness near the punctures in \S4.
The proof of amputation is given
in \S5, where we also review the proof that sewing of
minimal area metrics yield minimal area metrics [Zw2].
The new minimal area metrics which do not
correspond to quadratic differentials are given in \S6.

In an interesting paper K. Ranganathan [Ra] has independently
found a criterion for minimal area metrics in regions foliated by
a single homotopy class of geodesics. His result, as we will
see, also establishes flatness around the punctures. Since the lemma
presented in \S3 since does not assume a single foliation,
it may be useful to determine if regions of the surface
with multiple foliations are also flat.

\chapter{Preliminary Notions}

We will be concerned throughout with conformal metrics on
Riemann surfaces. A (conformal) metric $\rho(z) |dz|$ must
be invariantly defined, {\it i.e.} $\rho(z_1)|dz_1|$
$=\rho(z_2) |dz_2|$, where $z_1$ and $z_2$ are local
parameters. The conformal factor $\rho$ must be measurable
and non-negative everywhere.
The length of a curve $\gamma$, denoted as $l_\rho (\gamma)$
is given by $\int_\gamma \rho |dz|$, and the area of the
metric $\rho$ is given by $\int\int \rho^2 dxdy$. Both
length and area are invariantly defined.  The minimal
area problem we are considering is the following [Zw1]:

\noindent
$\underline{\hbox{Generalized Minimal Area Problem}}$.
Given a genus $g$  Riemann surface, with
$n\geq 0$ punctures ($n\geq 2$ for $g=0$) find the
metric of minimal area under the condition that the
length of any nontrivial homotopy closed curve be
greater than or equal to $2\pi$.
\medskip

In the introduction we have reviewed some properties of the
minimal area metric. Here we shall concentrate on other
aspects of this problem, of direct relevance for our
present work.

Whenever there are punctures the naive definition of area
is not adequate for the above problem. One must use the
``reduced area'' defined as [Zw1] (see also [St] \S3.2 )
$${\cal A}(\rho) = \lim_{r\rightarrow 0}
\Bigl( \int\hskip-6pt\int\limits_{\R (r)} \rho^2 dxdy
+ 2\pi n \ln r \Bigr) ,\eqn\redarea$$
where $n$ is the number of punctures and $\R (r )$ denotes the
Riemann surface obtained by excising the disks $|z_j|\leq r$
around the punctures. The $z_j$'s are arbitrary (but fixed)
local coordinates vanishing at the punctures. Under a
change of local coordinates the reduced area changes by a metric
independent constant, and therefore the notion of a metric
of minimal reduced area is independent of the choice of local
coordinates.

A metric $\rho$ is called admissible (for the minimal area problem)
if all nontrivial curves satisfy the length conditions. A curve
satisfying the length conditions will be called a ``good'' curve;
a curve that violates the length conditions will be called a
``bad'' curve. Let us emphasize that we will always be
dealing with simple closed curves, i.e. curves without
self-intersections.

Given two admissible metrics $\rho_0$ and $\rho_1$
the metric $\rho_t = (1-t)\rho_0 + t\rho_1$ is admissible for
all $t\in [0,1]$. The reduced area can be shown to be a strictly
convex functional: ${\cal A}(\rho_t) < (1-t) {\cal A}(\rho_0)
+ t{\cal A}(\rho_1)$ for $t \in (0,1 )$. This implies
that the metric of minimal reduced area is unique [Zw1] \foot{This
property is well known for the ordinary definition of area.
See, for example, K. Strebel [St].}

Let us now explain the relation
to extremal length. Consider, for simplicity, a genus $g$
surface $(g\geq 1)$ with no punctures (so that we can use
area rather than reduced area).
Given a family $F$ of curves on the surface,
the extremal length $\Lambda (F)$ is a conformal
invariant defined by
$$\Lambda (F) = \sup\limits_\rho {L(\rho )^2 \over A(\rho )},
\eqn\extremal$$
where the supremum is taken over all possible metrics,
$L(\rho ) \equiv \inf\limits_{\gamma \in F} l_\rho (\gamma )$,
and $A(\rho )$ is the area. Since the extremal length does
not change under $\rho \rightarrow c \rho$, with $c$ an arbitrary
positive constant, it is possible for every $\rho$ to demand
$L(\rho ) = 2\pi$, namely that all curves in $F$ be longer or
equal to $2\pi$. Then, it follows from \extremal\ that we
must try to make the area $A$ as small as possible. Thus
the metric of least area under the condition that all curves
in $F$ be longer than or equal to $2\pi$ (if it exists) will
give us the value of $\Lambda (F)$. Our minimal area problem
simply corresponds to the case when $F$ is the set of all nontrivial
curves on the surface.

\noindent
$\underline{\hbox{Remark}}$. The quantity $f = (2\pi)^2/A(\rho)$
for the minimal area
metric is expected to define an interesting function
$f :\overline \M_{g,0} \rightarrow {\bf R}$.
It is a function on moduli space since it is unique and requires no
choice of curves. At the compactification divisor it is expected to
go to zero since the area must diverge. Example: it is not
hard to show that for $g=1$, in terms of the modular parameter
$\tau$ in the usual fundamental domain $-1/2 \leq \tau < 1/2$,
$|\tau | \geq 1$ and $\hbox{Im}\, (\tau ) >0$; the function $f$
is simply $f = 1/\hbox{Im}\,(\tau )$. The torus with least minimal
area has $\tau = \exp (i\pi /3)$.

An important notion for us will be that of saturating geodesics.
These are the generalization, for our problem, of the closed
horizontal trajectories of Jenkins-Strebel quadratic differentials.
They are simply the geodesics of length $2\pi$, namely the geodesics
that saturate the length conditions of the minimal area problem.
We now show that under some conditions the saturating geodesics
give rise to a foliation.

\noindent
$\underline{\hbox{Lemma 2.1}}.\,$ Consider a punctured
Riemann surface $\R$ with a complete minimal area metric
$\rho_0$, smooth and nonvanishing in some neighborhood $N_i$
of each puncture $p_i$. Then there is a neighborhood $V_i$ around
each puncture $p_i$ which is foliated by saturating geodesics
homotopic to the puncture.

\noindent
$\underline{\hbox{Proof}}.\,$
Consider a puncture $p$ surrounded
by a neighborhood $N$ in which $\rho_0$ is smooth and non-vanishing.
Then let $\g_0\subset N$ be an embedded (simple) closed curve
homotopic to the puncture. Let $V$ be the neighborhood of the
puncture consisting of points which are at distance greater than
$3\pi$ from $\g_0$. Completeness of the metric guarantees the
existence of the neighborhood $V$: if no such neighborhood could
be found it would mean that every neighborhood of the puncture would
contain a point whose distance to a fixed point $x_0$ in the curve
$\g_0$ is smaller than $3\pi + l(\g_0)/2$, and then we would have a
sequence of points converging to the puncture whose distance to a
fixed point is bounded, in contradiction with the completeness
of $\rho_0$.

Then we claim that if $C$ is a nontrivial closed curve
which intersects $V$ but which is not
homotopic to $\g_0$, then $l_{\rho_0}(C)>3\pi$. To see this, notice
that $C$ must intersect $\g_0$, and the distance from $\g_0$ to
any point in $V$ is at least $3\pi$. Thus if there is a
saturating geodesic intersecting $V$ it must be homotopic to
the puncture.

We claim that there is a smooth curve through $q$ homotopic to $p$
of length $2\pi$.  We first observe that there must be a curve passing
through $q$ of length less than or equal to $2\pi + \e$, for
every $\e>0$ [Zw2]; if this were
not true, then there is some $\epsilon$ such that all nontrivial
closed curves through $q$ are longer than $2\pi + \epsilon$.
Consider then an $\e / 3$ neighborhood of $q$, and set the metric
equal to zero throughout this neighborhood. We claim the new metric
is admissible. If any curve becomes shorter than $2\pi$, it
is because its portion lying outside the neighborhood is smaller
than $2\pi$, but then the nontrivial open subcurve lying
outside the neighborhood could be made into a nontrivial closed curve by
joining its endpoints with $q$. In this way we would get a curve
of length smaller than $2\pi + 2\epsilon /3$ in contradiction with
the assumption that all nontrivial closed curves through $q$ had to be
longer than $2\pi + \epsilon$.

We consider now a family of curves
$\g_n$ passing through $q$ of lengths less than $2\pi + {1\over n}$.
All of these curves are
homotopic to the puncture $p$, since they all have lengths less than $3\pi$.
Now, extend $\rho_0 |_N$ to a smooth complete admissible metric $\rho_0$
on $\R$ and then lift this metric to a metric $\widetilde\rho_0$
on the universal
cover $\widetilde\R$.  Then, by the Hopf-Rinow theorem, we find a minimal
geodesic $\widetilde\g$ between the endpoints of a lift $\widetilde\g_n$
of $\g_n$.
Of course, this geodesic $\widetilde\g$
must have length less than $2\pi + {1\over n}$, for all $n$,
since it is minimizing.
Moreover, since $\widetilde\g$ will project to a curve homotopic to the
puncture and $\rho_0$ is admissible, the curve $\g$ must have length
at least $2\pi$.  We conclude that this curve has length exactly $2\pi$,
and thus, by the arguments above, $\g$ lives in $N$,
and is a $\rho_0$-geodesic.  Finally, $\g$ is smooth at $q$,
since otherwise we could cut any corner of $\g$ at $q$ and obtain a curve
of length less than $2\pi$, thus showing that $\rho_0$ is not admissible.

If there is more than one saturating geodesic through $q$ they
would have to intersect at some finite angle, since in a smooth
metric two geodesics going through the same point and having the
same first derivative must coincide completely.
If two saturating geodesics intersect
they would have to do so at least at two
points. Consider then two homotopic segments determined by
the intersection points. These must have equal lengths, and then,
a cut and paste argument shows that the length of the curves
could be reduced at the intersection points. Thus we have
established the uniqueness of the saturating
geodesic going through $q$ and the fact that homotopic geodesics
cannot intersect. This implies that the neighborhood $V$ will be
foliated completely by saturating geodesics homotopic to the
puncture. \hfill$\square$

\chapter{The Deformation of an Admissible Metric}

In this section we will establish a lemma that describes
deformations that can be made to admissible conformal metrics (in a
region covered by a finite number of foliations) so that the
deformed metric will remain admissible.  This lemma and its
corollary will be
used in \S4 to show that if a minimal area metric were not flat in
a neighborhood of each puncture,
then we could flatten it slightly and reduce its area,
contradicting the presumed minimality of the metric.
\medskip
\noindent
$\underline{\hbox{Lemma 3.1}}.\,$ Let $\R$ be a Riemann surface and
$\rho_0$ an admissible complete metric. Let $h$ be a smooth conformal
metric (admitting both positive and non-positive values) and suppose
that $h$ has compact support. Let $\G$ denote the set of
saturating geodesics passing through $\supp(h)$ and suppose that
\item{(i)} for every $\g_0\in\G$, the set
$\G_{[\g_0]}\subset\G$ of curves freely homotopic to $\g_0$ cover
$\supp(h)$, i.e.,
$\supp h\subset\bigcup\lm_{\g\in\G_{[\g_0]}}\gamma$, for
every $[\g_0 ]$ represented in $\G$,
\item{(ii)} the original metric $\rho_0$ is smooth ($C^\infty$)
and non vanishing
in a neighborhood of $\bigcup\lm_{\g\in\G}\g$,
\item{(iii)} $\G$ contains only a finite number of free homotopy
classes of curves,
\item{(iv)} there is a $\delta_0$ so that if
$\gamma$ is a nontrivial closed curve passing through $\supp(h)$ with
$l_{\rho_0}(\g)<2\pi+\delta_0$, then $\g$ is freely homotopic to a
curve in $\G$,
\item{(v)} $\int\lm_\g h |dz| \ge 0$ for each $\g\in\G$.
\medskip
\noindent
Then there is an $\hat{\e}$ and a constant $K$ independent of
$\hat{\e}$, such that for
$\e<\hat{\e}$,
the metric $\rho_\e$ defined by
$$
\rho_\e = \rho_0 + \e h + K\e^2\chi, \eqn\modifmetric
$$
is admissible. Here $\chi$ is a conformal metric which equals
$\rho_0$ on a $2\pi$ neighborhood of $\supp(h)$ and vanishes
elsewhere.

\medskip
\noindent
$\underline{\hbox{Proof}}.\,$ Let $\wh\rho_\e$ denote the truncated
$\rho_\e$ metric,
$$
\wh\rho_\e = \rho_0 + \e h.
$$
Note that the metric $\wh\rho_\e$ is well defined only if
$\e$ is small enough (otherwise the metric could become negative).
Since the support of $h$ is compact, over this support
$\inf (\rho_0)$ exists and
is different from zero (see (ii)), and
$\sup |h|$ exists and is different from zero or infinity.
One then checks that for $\e < \e_0 = \inf (\rho_0 ) /
\sup |h|$, the metric $\rho_\e$ is well defined.

We must check that all nontrivial closed curves on the surface
are still longer than or equal to $2\pi$ when the metric changes
from $\rho_0$ to $\rho_\e$, for $\e <\e_0$. We begin
by narrowing the space of curves that must be checked.
We now show that it is sufficient to consider only those curves
$\g_\e$ which satisfy the following four criteria:

\noindent
(a) $\g_\e$ must intersect $\supp (h)$; otherwise
its length can only increase when going from the $\rho_0$
to the $\rho_\e$ metric.

\noindent
(b) $\g_\e$ must be contained in supp$(\chi )$; if it
is not, then, because of (a) the part of the curve outside
supp($h$) must have length of at least $2\pi$. Since outside
supp($h$) we have $\rho_\e >\rho_0$, the curve $\g_\e$ has
$\rho_\e$-length at least $2\pi$.

\noindent
(c) The $\rho_0$ length of $\g_\e$ must be less or equal
to $4\pi$, namely $l_{\rho_0} (\g_\e ) \leq 4\pi$. Suppose
it is not, then $l_{\rho_0} (\g_\e ) > 4\pi$. Consider now
$$ l_{\wh\rho_\e} (\g_\e ) = \int_{\g_\e}
\wh\rho_\e |dz| = \int_{\g_\e} {\wh\rho_\e \over \rho_0}
\rho_0 |dz|,$$
for $\e < {1\over 2} \e_0$ one has
$${\wh\rho_\e \over \rho_0} > {1\over 2},\quad
\hbox{so that} \quad
l_{\wh\rho_\e} (\g_\e ) > {1\over 2} \int_{\g_\e}
\rho_0 |dz| > {1\over 2}\, 4\pi = 2\pi , $$
which shows that the curve $\g_\e$ will still be longer
than $2\pi$ in the modified metric. Thus indeed, we only
need to check those curves whose $\rho_0$ length is shorter
than $4\pi$.

\noindent
(d) The curve $\g_\e$ is freely homotopic to an element of $\G$.
Note that only if $l_{\wh\rho_\e} (\g_\e ) < 2\pi$
is there something to check, but in this case
$$l_{\rho_0} (\g_\e ) + \e \int_{\g_\e } h |dz| < 2\pi,$$
and so,
$$l_{\rho_0} (\g_\e ) < 2\pi + \e \int_{\g_\e } |h| |dz|.$$
The last term above can be bounded
$$ \e \int_{\g_\e } |h| |dz|=\e \int_{\g_\e }
|{h \over \rho_0}| \, \rho_0 \,|dz| \leq
\e \, {\sup |h| \over \inf (\rho_0)}
\, 4\pi \leq \delta_0 , $$
for $\e \leq {\delta_0 \over 4\pi} \e_0$.
We therefore have $l_{\rho_0} (\g_\e ) < 2\pi + \delta_0$
and the desired result is a consequence of hypothesis (iv).
In summary we need only discuss curves contained in the support
of $\chi$, and homotopic to elements of $\G$.
\medskip
We now begin our analysis of the relevant curves.
Since $\g_\e$ is
freely homotopic to an element of $\G$ and such elements cover
$\supp(h)$, there is a curve
$\g_0\in\G$ which is homotopic to $\g_\e$, intersecting
$\g_\e$ in at least two distinct
points, say $p_1$ and $p_2$. To see this, observe that by (a) above,
$\g_\e$ must intersect the interior of supp($h$), and so must
intersect some $\g_0 \in \Gamma$ which passes through the interior
of supp($h$). Then, if $\g_0 \cap \g_\e$ consists of only a single
point, by replacing $\g_0$ with a nearby curve of the foliation near
$\g_0$, we may assume that $\g_0 \cap \g_\e$ contains two points.

Let $A_0$ and $A_\e$ denote the arcs of $\g_0$ and $\g_\e$,
respectively, which pass through $p_1$ and $p_2$ and are homotopic
rel $\{p_1,p_2\}$ (see Fig.~1). Our goal is to bound from below the
$\wh\rho_\e$-length of $A_\e$ in terms of the $\wh\rho_\e$-length
of $A_0$, say $l_{\wh\rho_\e}(A_\e)>l_{\wh\rho_\e}(A_0)-K_3\e^2$,
so that we can eventually bound the $\wh\rho_\e$-length of $\g_\e$
in terms of the $\wh\rho_\e$-length of $\g_0$, which by hypothesis
exceeds $2\pi$.

To this end, set $\wh\rho_t=\rho_0+th$ and let $A_t$ denote the
$\wh\rho_t$-geodesic arc connecting $p_1$ and $p_2$; here
$A_0\subset\g_0$. We wish to consider $l_{\wh\rho_t}(A_t)$ as a
function of $t$, and, in particular, we wish to differentiate this
function in $t$. To begin this discussion, we first need to
establish how $A_t$ varies with $t$. For this we recall the theorem
of Eells and Lemaire [EL; \S4] that for each pair of points
$p_1$ and $p_2$ connected by a geodesic arc $A_0$, there is an
$\e_0$ (depending on the pair $(p_1,p_2)$ as well as $h$ and
$\rho_0$) so that for $t<\e_0$, we find a unique family of
$\wh\rho_t$-geodesic arcs $A_t$ connecting $p_1$ and $p_2$ as long as
there are no non-trivial $\rho_0$-Jacobi fields along $A_0$
vanishing at the endpoints. Of course,
there are no such Jacobi fields along $A_0$, because if
there were, then there would be a non-trivial Jacobi field along
$\g_0$, and we could use such a field to find a curve $\g^*_0$ of
lower length than $\g_0$, contradicting the admissibility of
$\rho_0$ (see, for example [Sp] Vol.4 Ch. 8 or [Hi] Ch. 10, Thm 11).
Thus, we can find such an $\e_0$ as described above;
however, in order to argue for all extremal arcs $A_\e$, we need an
$\e_0$ that does not depend on any particular $A_0$.
To find such an $\e_0$, we use the compactness of the space of
endpoints $\{(p_1,p_2)\}$ as well as the smooth dependence of
solutions of the Jacobi equation $J(V)=0$ upon the data. More
precisely, let $N\subset\R$ denote the submanifold of $\R$
consisting of all points within
$\supp( \chi )$. Here the completeness of $\rho_0$ implies that $N$ is a
compact submanifold of $\R$. Form the compact product manifold
$N\x N$ and consider within $N\x N$ the submanifold $X$
consisting of pairs of points $(p_1,p_2)$ with the property that
$p_1$ and $p_2$ lie along a $\rho_0$-geodesic belonging to $\G$,
and that $d_{\rho_0}(p_1,p_2)\le 2\pi$.
Now for each of the finite number of homotopy classes represented
in $\G$, there is a compact family of $\rho_0$-geodesics
representing that homotopy class, and so we can conclude that $X$
is compact.

Finally, we want to form a compact space representing all the arcs
of curves in $\Gamma$; to do this we need to account for there
being two arcs in a curve in $\Gamma$ which have endpoints represented
by $(p_1,p_2) \in X$. To this end, consider for each of the finitely
many homotopy classes in $\Gamma$ a given fixed orientation, then
we form the disjoint union $X_0 = X \sqcup X$ so that the point
$(p_1,p_2)$ in the first copy will correspond to the arc from
$p_1$ to $p_2$ determined by the orientation of the underlying curve
$\g \in \Gamma$ and $(p_1,p_2)$ in the second copy will correspond
to the arc from $p_1$ to $p_2$ determined by the opposite orientation
of the underlying curve $\g\in \Gamma$. The set $X_0$ is compact,
as desired.

Now, for each curve $\g\in\G$, we can consider the
lift $\wh\g$ to a unit speed periodic extension
$\wh\g: \BR\to\R$ where $\wh\g([0,2\pi))$ covers the embedded curve
$\g$ exactly once. Then, because there are no non-trivial Jacobi
fields on $\g$, there is a $\delta(\g)$ so that there are no
non-trivial solutions to the Jacobi equation with vanishing
boundary values for $\wh\g$ on the interval $[0,l]$ for
$l<2\pi+\delta(\g)$. Since $\G$ is compact and the solutions to the
Jacobi equation vary smoothly with the data, we conclude that there
is a $\delta>0$ so that for all $\g\in\G$, there are no non-trivial
solutions to the Jacobi equation with vanishing boundary values for
$\wh\g$ on the interval $[0,l]$ for $l<2\pi+\delta$.

{}From this uniform non-degeneracy of the Jacobi operator on boundary
value problems with data from $X_0$ and the smoothness of $\rho_0$
near $\bigcup\lm_{\g\in\G}\g$, the proof of the Eells-Lemaire
result implies that for every pair of points $x_0=(p_1,p_2)\in X_0$
there is a $T^*(x_0)$ and a neighborhood $\N$ of $x_0$ in $X_0$ so
that for $|t|<T^*(x_0)$ and $x\in\N$, there is a unique and
differentiable family of arcs $A_t$ which are geodesic in the metric
$\wh\rho_t$ and have endpoints given by $x\in\N\subset X_0$. Using
the compactness of $X_0$, we find a $T^*$ so that for $|t|<T^*$ and any
pair of points $p_1$, $p_2\in\g\in\G$, there is a unique and
differentiable family of $\wh\rho_t$-geodesic arcs $A_t$ connecting
$p_1$ and $p_2$. Also, the compactness of the family of arcs obtained
for $|t|\le T^*/2$ together with the result of the last
paragraph guarantees a $T_0$ with the same property as $T^*$ above but
with the additional property that there do not exist any non-trivial
$\wh\rho_t$-Jacobi fields along any of the arcs $A_t$ for
$|t|\le T_0/2$.

We are interested in estimating the derivatives of the function
$l_{\wh\rho_t}(A_t)$, and it is here that the final property of the
arcs $A_t$ described in the previous paragraph becomes important.
We will need to know
$$
{\p^2\over\p t^2}\biggm|_{t=c} l_{\wh\rho_c} (A_t)\eqn\threeone$$
for all values $c$ with $|c|<\e<T_0/2$. It is convenient to introduce
here the rectangular parameter space $(s,t)$ with $s \in [0,1]$ and
$t\in (-\e ,\e )$. We think of $A_t(s)$ of maps from the parameter
space to the manifold, giving us, for constant $t$, the arc $A_t$.
In addition, $A_t(0) = p_1$ and $A_t(1) = p_2$ for all $t\in (-\e ,\e )$.
The second variation formula
for length provides that \threeone\ can be bounded in terms of
$\| W_c (s) \|_\infty$,  where
$W_c (s) = {\p\over\p t}\bigm|_{t=c}A_t(s) = A_{c*} {\partial
\over \partial t}$
is the tangent vector in the surface showing us how the geodesic
segments move as we change $t$, and in terms of
$\| \nabla_{\T_c}^{\wh\rho_c}\, W_c \|_\infty$, where
$\T_c = A_{c*}{\p\over\p s}$ is the tangent vector to the geodesic
arcs. We will now estimate those quantities.

To begin we consider a particular family $A_t$ with endpoints
$(p_1,p_2)\in X_0$ which is geodesic in the metric $\wh\rho_t$, and
using a coordinate system $(x_1,x_2)$ in a neighborhood of the
$\wh\rho_c$-geodesic $A_c$, we compute the Christoffel symbols
$\G(t)=\G^k_{ij}(t)$ and find
that for the conformal metric $\wh\rho_t$, we may define a symbol
$\Psi^k_{ij}(t)$ so that
$$
\G^k_{ij}(t) = \G^k_{ij}(c) + \Psi^k_{ij}(t)
$$
where $\Psi^k_{ij}(c)=0$. Moreover, since we can take $\wh\rho_t$
non-vanishing in a neighborhood of $\bigcup\lm_{\g\in\G}\g$ which
contains all of our arcs $A_t$, $|t| <T_0/2$,
and $h$ is smooth, we
see that $\Psi^k_{ij}(t)$ is not only differentiable in $t$ but is
also smooth in $(x_1,x_2)$, for $|t|<T_0/2$. Then the $\wh\rho_t$
geodesic arc $A_t$ satisfies
$$
(A_t)^k_{,ss} + \G^k_{ij}(t) (A_t)^i_{,s} (A_t)^j_{,s} = 0,
\eqn\threetwo
$$
where each curve $A_t: [0,1]\to\R$ is parametrized proportionally
to arc length.

We rewrite \threetwo\ as
$$
(A_t)^k_{,ss} + \G^k_{ij}(c) (A_t)^i_{,s} (A_t)^j_{,s} =
\Psi^k_{ij}(t) (A_t)^i_{,s} (A_t)^j_{,s}.\eqn\threethree
$$
Next we use $W_c(s) ={\p\over\p t}\bigm|_{t=c}A_t(s)$ and then
differentiate \threethree\ in the variable $t$ at $t=c$.
It is straightforward (and classical) that the left hand side of
\threethree\ differentiates to give the Jacobi operator
$\nabla_{\T}\nabla_{\T}
W + R(W, \T) \T$ where
we take the covariant derivatives and curvatures with respect to
the metric at $t=c$, that is $\wh\rho_c$. To evaluate the right
hand side of \threethree, introduce $\wh\rho_c$-Fermi coordinates along
the arc $A_c$ so that
$A_c(s)= (A_c^1 , A_c^2 ) = (l_{\wh\rho_c}(A_c)s,0)$ and recall
that $\Psi^k_{ij}(c)=0$. Then we find that
$$
\eqalign{
-{d\over dt}\biggm|_{t=c} [\Psi^k_{ij}(t) (A_t)^i_{,s}
(A^j_t)_{,s} &= -\left[{d\over dt}\biggm|_{t=c} \Psi^k_{11}(t)\right]
[l_{\wh\rho_c}(A_c)]^2\cr
&\equiv \Psi^k (A_c(s)) \, [l_{\wh\rho_c}(A_c)]^2 , \cr}\eqn\normalize
$$
where the last expression indicates a local function defined along
the arc $A_c(s)$. We are left with the vector equation
$$
J(W) = \nabla_\T \nabla_\T W + R\left(W,
\T \right)\T = \vec\Psi(A_c(s))\, [l_{\wh\rho_c}(A_c)]^2.\eqn\threefour
$$
It is now convenient to introduce the vector field
$T_t = \T_t/l_{\wh\rho_t}(A_t)$, with unit normalization:
$<T_t , T_t >^{1/2}$ $= <\T_t , \T_t >^{1/2} \hskip-2pt /
[l_{\wh\rho_c}(A_c)] = 1$
(recall $\T_t = A_{t*} {\partial \over \partial s}$, and $s\in [0,1]$).
Therefore Eqn. \threefour\ is rewritten as:
$$
\nabla_T \nabla_T W + R\left(W,
T \right) T = \vec\Psi(A_c(s)) .\eqn\getitright
$$
Now, by our choice of $|c|<T_0/2$, we know that there are no Jacobi
fields along $A_c$. Thus the general theory of ordinary
differential equations (cf [H], Theorem~XII.3.1) ensures a bound
$K_{p_1,p_2;c}$ so that the solution $W=W_{p_1,p_2;c}$ of \getitright\
with boundary conditions
$W_{p_1,p_2;c}(p) = W_{p_1,p_2;c}(p_2)=0$ satisfies
$$
\{ \|W_{p_1,p_2;c}\|_{\infty} \, ; \,
\| \nabla_{T_c} \, W_{p_1,p_2;c}\|_{\infty}\}
< K_{p_1,p_2;c}\int\lm^\theta_0
\|\vec\Psi\| ds.\eqn\threefive
$$
Now, the bounds $K_{p_1,p_2;c}$ are determined by the coefficients
of the homogeneous equation associated to \threefive, so after possibly
replacing $K_{p_1,p_2;c}$ with $4\pi K_{p_1,p_2;c}$ we conclude
that for all pairs of points $x=(q_1,q_2)\in X_0$ near
$(p_1,p_2)\in X_0$ and $t$ near $c$, then
$$
\{ \|W_{x;t}\|_{\infty} \,;\,
\| \nabla_{T_t}^{\widehat\rho_t} \,
W_{x;t}\|_{\infty} \} \le K_{p_1,p_2;c} \|\Psi\|_{\infty}.
$$
Upon choosing a finite cover of $X_0\x[-T_0/2,T_0/2]$ by neighborhoods of
$(x_i,c_i)$
$\in X_0 \times [-T_0/2,T_0/2]$, we find the bound
$$
\{ \|W_t\|_{\infty} \,;\,
\| \nabla_{T_t}^{\widehat\rho_t} \, W_t \|_{\infty} \}
\le K_0 ,\eqn\threesix
$$
for all solutions to \threefour\ along
$\wh\rho_t$-geodesic arcs $A_t$, and for all $t\in[-T_0/2,T_0/2]$.

We turn finally to using our estimate \threesix\ to bound
$l_{\wh\rho_\e}(A_\e)$ from below for $\e$ sufficiently small,
where $A_\e$ is the portion of the minimal $\wh\rho_\e$-geodesic
$\g_\e$ constructed at the outset of the argument. Because
$l_{\wh\rho_t}(A_t)$ is differentiable in $t$, we have the
expansion
$$
l_{\wh\rho_\e}(A_\e) = l_{\rho_0}(A_0) + \e{d\over dt}\bigm|_{t=0}
[l_{\wh\rho_t}(A_t)] + {\e^2\over 2}{d^2\over dt^2}\bigm|_{t=c}
[l_{\wh\rho_t}(A_t)]\eqn\threeseven
$$
for some $c\in(0,\e)$. In order to facilitate the evaluation of
the derivatives it is convenient to introduce the following
function of two variables
$$l(t,w) = l_{\widehat\rho_t} (A_w) = \int_{A_w} \widehat\rho_t |dz|,
\eqn\twovar$$
where the first variable $t \in [-T_0/2 ,T_0 /2]$ labels the
different metrics, and the second variable $w \in [-T_0/2,T_0 /2]$
the different arcs. Clearly
we have $l_{\widehat\rho_t} (A_t) = l(t,t)$ and therefore we simply have
that the terms entering equation \threeseven\ are given by
$$
{d\over dt}\, l(t,t) = {\partial\over \partial t} \, l(t,w)
\biggm|_{w=t} + \, {\partial\over \partial w} \, l(t,w)
\biggm|_{w=t} ,\eqn\sssww$$
and,
$$
{d^2\over dt^2}\, l(t,t) ={\partial^2\over \partial t^2} \, l(t,w)
\biggm|_{w=t} +\,\,  2 \,{\partial\over \partial t}
{\partial\over \partial w} \,l(t,w) \biggm|_{w=t}
+\, {\partial^2\over \partial w^2} \,l(t,w) \biggm|_{w=t},
\eqn\takederiv$$
where we have used the fact that the partial derivatives
commute. Consider now the second term in the right hand side of \sssww .
It corresponds to a first length variation for a family of curves;
since the derivative is evaluated at $w=t$ and the corresponding
curve $A_t$ is a geodesic in the metric
$\widehat\rho_t$, this term vanishes identically for all $t$:
$$ {\partial\over \partial w} \,l(t,w)
\biggm|_{w=t} \equiv  0, \quad \hbox{for} \,\, t \in [-\e ,\e ].
\eqn\vanishesident$$
Taking a $t$ derivative of this equation
$$ {d\over dt} \,\biggl( \,{\partial\over \partial w} \,l(t,w)
\biggm|_{w=t} \,\biggr) \, =  0, \quad \hbox{for} \,\, t \in [-\e ,\e ].
\eqn\vanishesidentic$$
we obtain the following relation
$${\partial\over \partial t}
{\partial\over \partial w} \,l(t,w) \biggm|_{w=t} =-
\, {\partial^2\over \partial w^2}
\,l(t,w) \biggm|_{w=t}.\eqn\simplify$$
Using \vanishesident\ and \simplify , equations \sssww\ and
\takederiv\ reduce to
$$\eqalign{
{d\over dt}\, l(t,t) &= {\partial\over \partial t} \,l(t,w)
\biggm|_{w=t} = \int_{A_t} h |dz| ,\cr
{d^2\over dt^2}\, l(t,t) &={\partial^2\over \partial t^2} \,l(t,w)
\biggm|_{w=t}- \, {\partial^2\over \partial w^2} \, l(t,w) \biggm|_{w=t}
= - \,{\partial^2\over \partial w^2} \, l(t,w) \biggm|_{w=t}, \cr}
\eqn\derivalmost$$
where we used \twovar\ with $\widehat\rho_t =\rho_0 + th$, so that
${\partial^2 \over \partial t^2} \wh\rho_t = 0$.
We now claim that
$$\biggm| \,
{\partial^2\over \partial w^2} \,l(t,w) \biggm|_{w=t}\, \biggm|
\, < \, 2K,\eqn\bounding$$
where $K$ is a bound independent of the endpoints of the geodesic
and of the value of $t \in (-\e ,\e )$. This follows because the
above is precisely the second length variation for a family of
curves, evaluated for a base curve which is a geodesic. Using
the standard formula (see Hicks [Hi] Chapter 10, Corollary, p. 151)
$${\partial^2\over \partial w^2} \,l(t,w) \biggm|_{w=t}
= \int_0^{l(A_t)} d\sigma \,\bigl( <R(W, T ) W,T >
+ <\nabla_{T} W , \nabla_{T} W > - (T <W,T >)^2 \,\bigr),$$
we see that the boundedness of curvature, together with the uniform
bounds of equation \threesix\ guarantee the bound in \bounding .

Using equations \derivalmost\ and \bounding\ we find that equation
\threeseven\ gives
$$
l_{\wh\rho_\e} (A_\e) > l_{\rho_0}(A_0)
+ \int_{A_0} h|dz| - K\e^2 .\eqn\threeeleven
$$
This inequality
refers to an arc $A_0\subset\g_0$ running from
$p_1$ to $p_2$ along $\g_0$. There is a similar inequality
for the other subarc $A'_0\subset\g_0$ continuing
along $\g_0$ from $p_2$ back to $p_1$ (see Fig.~1).
Adding the two we find
$$
l_{\wh\rho_\e}(\g_\e)  > l_{\rho_0}(\g_0)
+ \int_{\g_0} h|dz| - 2K\e^2 \geq  2\pi - 2K\e^2,
\eqn\whouw
$$
using hypothesis $(v)$ and the fact that $\g_0$ is a saturating
geodesic.

Thus, in the metric $\rho_\e=\wh\rho_\e+{1\over\pi}K\e^2\chi$, for
$\e$ sufficiently small, the shortest curve $\g_\e$ will have
length
$$
\eqalign{l_{\rho_\e}(\g_\e)  &= l_{\wh\rho_\e} (\g_\e) + {1\over\pi}
K\e^2 l_{\rho_0}(\g_\e)\cr
&> 2\pi - 2K\e^2 + {1\over\pi} K\e^2 (2\pi) = 2\pi , \cr}
$$
where we have used that the admissibility of $\rho_0$ forces
$l_{\rho_0}(\g_\e)\ge 2\pi$ and that $\g_\e\subset\supp\chi$. Thus
$\rho_\e$ is admissible. This concludes our proof
of the lemma. \hfill $\square$.
\medskip
Consider now a surface with a metric $\rho_0$ of finite (reduced) area
$A(\rho_0)$. The (reduced) area $A(\rho_\epsilon)$ of the
surface with the $\rho_\epsilon$ metric defined in \modifmetric\
is given by
$$\eqalign{
A(\rho_\e ) &= A(\rho_0) + 2\e \int h\rho_0 dxdy
+\e^2\int (2K\chi \rho_0 + h^2) dxdy \cr
 &\,\,\,\,\, + 2\e^3 K \int h\chi dxdy + \e^4 K^2 \int \chi^2 dxdy. \cr}
\eqn\areapert
$$
This result is obvious when there are no punctures and the $A$'s
denote (ordinary) area. It holds when there
are punctures, and we use reduced area, because the metric $\rho_0$ is
complete and therefore the (compact) supports of $h$ and $\chi$
avoid some neighborhoods of the punctures.
The $\O (\e )$ and $\O (\e^3 )$ terms
above are necessarily finite, since both $\rho_0$ and $h$ are
smooth over the support of $h$. The $\O (\e^2 )$ and $\O (\e^4 )$
terms are also finite since $\rho_0$ must be square integrable
over the support of $\chi$ (otherwise its (reduced) area
would be infinite). Then, if the following inequality holds
$$
\int h \rho_0 dxdy < 0,\eqn\thecondition
$$
we have that $A(\rho_\e ) < A(\rho_0)$ for sufficiently
small $\e$. If $\rho_0$ and $h$ satisfy the conditions of
Lemma~3.1, the metric $\rho_\e$ is admissible for sufficiently small
$\e$, and then \thecondition\ implies that $\rho_0$ cannot
be of minimal area. We have therefore established
the following corollary:

\noindent
$\underline{\hbox{Corollary 3.2}}.\,$ Given a metric $\rho_0$
of finite (reduced) area and a smooth variation $h$ satisfying
the conditions of Lemma~3.1, then $\int h\rho_0 dxdy <0$ implies
that the metric $\rho_0$ is not of minimal area.

\noindent
$\underline{\hbox{Remark}}.\,$ Lemma~3.1 together with
Corollary~3.2 give a partial local converse to Beurling's
criterion [Ah]. In Beurling's  criterion a sufficient (but not
necessary) condition for extremal metrics is given. It
involves showing that for any $h$ satisfying condition
$(v)$ one must always be able to prove that the right
hand side of \thecondition\ is greater or equal to zero.
Our result is a partial converse because we have shown that for
suitable $h$ and $\rho$ Beurling's condition is necessary.
It is a local result because it tests locally whether the area
of a metric can be reduced.

\chapter{Neighborhoods of the Punctures are Flat}

In this section, we discuss the geometry of the minimal area metric
in neighborhoods of the punctures of our punctured Riemann surface
$(\R; p_1,\dots,p_n)$. In particular, we will show that these
neighborhoods are isometric to flat cylinders of circumference
$2\pi$, foliated by a parallel family of geodesics of length
$2\pi$.

The idea will be to apply Corollary~3.2 to the case of a
complete minimal area metric which is
smooth in the neighborhood of a puncture.
We show that if the metric is not flat, then it admits a deformation
$h$ (as in Lemma~3.1) which lowers its area.
\medskip
\noindent
$\underline{\hbox{Theorem 4.1}}$. Let $\rho_0$ be a complete minimal area
metric on a Riemann surface $\R$ with punctures $p_1,\dots ,p_n$.
Suppose that each puncture $p_i$ is contained in a neighborhood
$N_i$ in which $\rho_0$ is smooth and non-vanishing. Then there are
neighborhoods $V_i$ around $p_i$ in which $\rho_0$ is isometric to
a flat semiinfinite cylinder of circumference $2\pi$.
\medskip
\noindent
$\underline{\hbox{Proof}}.\,$  Consider a puncture $p$ surrounded
by a neighborhood $N$ in which $\rho_0$ is smooth and non-vanishing.
Lemma 2.1. implies that there is a neighborhood $V$
foliated by geodesics of length $2\pi$, each homotopic to the
puncture $p$. We claim that the leaves of the foliation are parallel,
in the sense that if $\g_1(s)$ and $\g_1(\s)$ are two distinct leaves
of the foliation, then
$\min\lm_\s d_{\rho_0}(\g_1(s), \g_2(\s))=\min\lm_{s,\s}
d_{\rho_0}(\g_1(s), \g_2(\s))$. To argue for this and the
conclusion of the theorem, we work on the infinitesimal level, and
consider the leaves of the foliation near a leaf $\g(0)$ as a
variation of curves $\g(t)$. In particular, near $\g(0)$, we
parametrize each leaf $\g(t)$ by arclength ``$s$'' so that
$\g(t)=\g(s,t)$ and we claim that the variation field
${\p\over\p t}\g(s,t)=V(s)$ is constant along $\g(0,s)$.

To see this consider a point $q\in\g(s,0)$ at which
${\p\over\p s}\|V(s)\|_{\rho_0}=\alpha\neq 0$; we will take
$\alpha<0$.

We seek a more convenient parametrization of $\g(s,t)$. To this
end, consider the vector field $X_0(s,t)$ of $\rho_0$-unit vectors
which are perpendicular to the curves $\g(t)$. Since the $\rho_0$
metric is smooth, the vector field $X(s,t)$ is smooth and hence
integrable; let the integral curve normal to $\g(\cd,0)$ at
$\g(s,0)$ be denoted by $N(s)$. We now parametrize $\g(s,t)$ so
that $N(s)$ intersects $\g(\cd,t)$ at $\g(s,t)$. Then,
by Gauss' lemma, $\g(s,t)$ is still a
parametrization by arclength for each fixed $t$. We normalize the
parametrization by taking $q=\g(0,0)$ and assuming that the curve $N(0)$
is parametrized by arclength by the variable $t$ (see Fig.~2).
Considering $\g (s,t)$ as a map from $[ 0,2\pi ] \times (-\e ,\e )$
to $\R$ we set $X(s,t) = \g_* {\partial \over \partial t} \bigm|_{s,t}$;
then $V(s) = X(s,0)$. While $\| X(0,t) \| = 1$ because the curve
$N(0)$ is parametrized by arclength, $\|X(s,t) \|$ is not a priori
of unit length. At the point $q$ we have assumed
${\partial \over \partial s} \| V(s) \| <0$ (at $s=0$).

Let $\G_-$ denote the subarc of $N(-\eta)$ with endpoints
$\g(-\eta,0)$ and $\g(-\eta,\kappa)$ and let $\G_+$ denote the
subarc of $N(\eta)$ with endpoints $\g(\eta,0)$ and
$\g(\eta,\kappa)$. Roughly our plan is to exploit the fact
that the length of $\G_-$ is bigger than that of $\G_+$
since the norm of the variation field has negative derivative.
Let $R_-$ denote the intersection of the
$\rho_0$-neighborhood of $\G_-$ of size $\delta$
with $\bigcup\lm_{0\leq t\leq \kappa} \g (\cdot , t)$.
We observe that
since the leaves of the foliation are perpendicular to $\G_-$, then
$R_-$ can also be described as the union of arcs of the foliation
of $\rho_0$-length $2\delta$ centered along $\G_-$. We similarly
define $R_+$ as the intersection of the
$\rho_0$-neighborhood of $\G_+$ of size $\delta$ with
$\bigcup\lm_{0\leq t\leq \kappa} \g (\cdot , t)$ (see Fig.~2).

Let $k$ and $k'$ be  smooth positive functions supported on the intervals
$(-\delta,\delta)$ and $(0, \kappa $) respectively.
Consider then the smooth positive function
$K :{\bf R}^2 \rightarrow {\bf R}$, defined as $K(s,t) = k(s)k'(t)$.
This function is compactly supported in a rectangle.
We use $K$ to define the smooth (metric) $h$ that we need;
$h$ will be supported on $R_-\cup R_+$.
For any point $q$ with coordinates $( s(q),t(q))$
we define $h$ by
$$\eqalign{
h(q) &= - \rho_0\, K(\, s(q)+\eta \, , \, t(q)\, )
\quad \hbox{for}\,\, q\in R_- , \cr
h(q) &= \, \, \rho_0\, K(\, s(q) - \eta \, , \, t(q) \, )
\quad  \, \hbox{for} \,\, q\in R_+ ,\cr
h(q) &= {} 0, \quad \hbox{elsewhere} .\cr}\eqn\defperturb
$$
Consider the metric $\wh\rho_\e=\rho_0 +\e h$.
Conditions $(i)-(iv)$
of Lemma~3.1 are satisfied so we now verify
condition $(v)$. For any saturating geodesic
$\g (t)$ one should have $\int_{\g(t)} h |dz| \geq 0$;
since $\rho_0 |dz|$ is the length element $ds$ we have
$$\eqalign{
\int_{\g(t)} h |dz| = \int_{\g(t)} {h\over \rho_0} ds \,
&= \int_{-\eta -\delta}^{-\eta +\delta} {h(s,t)\over \rho_0} ds
+ \int_{\eta -\delta}^{\eta +\delta} {h(s,t)\over \rho_0} ds ,\cr
{}&= -\int_{-\eta -\delta}^{-\eta +\delta} K(s+\eta ,t) ds
+ \int_{\eta -\delta}^{\eta +\delta} K(s-\eta ,t) ds ,\cr
{}&= - \int_{ -\delta}^{\delta} K(s,t) ds
+ \int_{-\delta}^{\delta} K(s,t) ds = 0, \cr}\eqn\preservelen$$
thus our deformation precisely preserves the length of the
saturating geodesics. Since $h$ satisfies all the conditions
of Lemma~3.1 we can now apply Corollary~3.1 to show that the
metric cannot be of minimal area. We must simply verify that the
quantity indicated in \thecondition\
is negative:
$$
\int\hskip-12pt\int\limits_{R_-\cup R_+} h\rho_0 dxdy =
\int\hskip-12pt\int\limits_{R_-\cup R_+} {h\over \rho_0}
 dA(\rho_0)< 0. \eqn\estim
$$
The left hand side of the inequality is given explicitly by
$$
\int\lm^\kappa_0\int\lm^{-\eta+\delta}_{-\eta-\delta}
{h(s,t) \over \rho_0}
\|\g_* {\p\over\p s}\|_{\rho_0} \| \g_*{\p\over\p t}\|_{\rho_0} \, dsdt
+ \int\lm^\kappa_0\int\lm^{\eta+\delta}_{\eta-\delta}
{h(s,t)\over \rho_0}
\|\g_* {\p\over\p s}\|_{\rho_0} \|\g_*{\p\over\p t}\|_{\rho_0}\, dsdt ,
\eqn\estimi
$$
since the $(s,t)$ coordinate system is orthogonal. Using
$\|\g_* {\p\over\p s}\|_{\rho_0}=1$  (parametrization is
by arclength) and $\|\g_* {\p\over\p t}\|_{\rho_0}= \| X(s,t) \|$,
we find that \estimi\ becomes
$$
-\int\lm^\kappa_0\int\lm^{-\eta+\delta}_{-\eta-\delta}
K(s+\eta ,t )\, \|X(s,t)\| \, dsdt + \int\lm^\kappa_0
\int\lm^{\eta+\delta}_{\eta-\delta} K(s-\eta , t)
\, \|X(s,t)\| \, dsdt, \eqn\estimii
$$
and finally, shifting the domains of integration we obtain
$$
-\int\lm^\kappa_0\int\lm^{\delta}_{-\delta}
K(s,t)\, g(s,t) \, dsdt , \quad
\hbox{where} \quad
g(s,t) = \|X(-\eta +s,t)\| - \|X(\eta + s,t)\| .\eqn\estimiii
$$
Since $g(0,0) =  \| V(-\eta ) \| -\| V(\eta ) \| $, and,
by assumption, ${\partial\over\partial s} \| V(s) \| < 0$ for $s=0$,
for sufficiently small $\eta$ one must have
$g(0,0)>0$. Since $g$ is continuous, it follows that
it is positive throughout the region of integration for
sufficiently small $\delta$ and $\kappa$. It is then
clear that the above expression is strictly negative,
as we wanted to show. This proves that
${\partial \over \partial s} \| V(s) \| = 0$, at
$q \in \g (0)$ (corresponding to $s=0$). Since this point is not a
special point, it follows that
${\partial \over \partial s} \| V(s) \| = 0$,
for all points in the geodesic $\g (0)$. Since the geodesic $\g (0)$
is not a special geodesic, we have
${\partial \over \partial s} \| X (s,t) \| = 0$ along every
leaf of the foliation. Since $\| X(0,t)\| = 1$, we have shown that
$\| X(s,t) \| = \| \g_* {\partial \over \partial t} \| = 1$.

Finally, we consider the map $\g(s,t)$ as a map from a flat
cylinder $C$ with locally Euclidean metric $\ov\rho=ds^2+dt^2$ to
the neighborhood $V$ of the puncture with metric $\rho_0$. The map
$\g(s,t)$ is a (local) diffeomorphism, and so we consider the
pullback metric $\g(s,t)^*\rho_0$ as a smooth metric on the
cylinder $C$. In comparing the pullback metric $\g(s,t)^*\rho_0$
with the flat metric $\ov\rho$, we first notice that both
$\g_* {\partial\over \partial s}$ and
$\g_* {\partial\over \partial t}$ are of unit
$\rho_0$-length and orthogonal for all $s$ and $t$: we conclude
that $\g(s,t)^*\rho_0=\ov\rho$ and so $(V,\rho_0)$ is isometric to
a flat annulus. This concludes the proof of Theorem~4.1.\hfill
$\square$

\medskip
\noindent
$\underline{\hbox{Remark}}.\,$ One often shows that for extremal
length problems, the neighborhoods in which the metric is smooth
must be flat (see [St] for example), and it is common to use the
length-area method in such arguments. Of course, the length-area
method requires an analysis along a neighborhood of an entire
geodesic; in the above argument, however, the analysis was
restricted to a neighborhood of a point, after requiring some
smoothness and finiteness properties of the neighborhood of the
geodesic.

\chapter{Amputation and Plumbing of Minimal Area Surfaces}

The purpose of the present section is to give a proof of
the amputation property of generalized minimal area metrics
on punctured surfaces. This will be done in section \S5.1 .
In \S5.2 we will discuss, for completeness of exposition,
the plumbing of minimal area metrics. Expanding the discussion
of [Zw2] we show that the plumbing of surfaces with metrics
of minimal area, using the canonical coordinates induced by
these metrics, induces on the resulting surface a  minimal area
metric.

Before beginning our exposition let us establish a useful
lower bound on lengths of nontrivial closed curves.
Consider a Riemann surface with an
admissible metric and a ring domain $F_i$ on the surface isometric to a
flat cylinder (finite or semiinfinite) of circumference $2\pi$.
This ring is foliated by saturating geodesics homotopic to its
core curve. Denote one of the boundaries of the ring domain by
$\C_0$, and let the curve $\C_\delta$ denote the saturating geodesic
at a distance $\delta$ away from $\C_0$. Let $\C_h$ denote the other
boundary of the ring domain; $h$ is called the height of the cylinder
($h=\infty$, if $F_i$ is a semiinfinite cylinder).
We say that a nontrivial
closed curve $\g$ penetrates $F_i$ a distance $\delta$
($\delta < h$) if one can
find two points $p_1,p_2 \in (\g \cap \C_0)$ and an open subcurve
$\g'$ of $\g$ with endpoints $p_1$ and $p_2$, fully contained in $F_i$,
such that $\g'\cap \C_\delta \not=0$, and $\g'\cap \C_\eta =0$,
for $h > \eta > \delta$.

We say that a nontrivial closed curve $\g$ extends a distance $\delta$
in a ring domain $F_i$ (as above) if the curve is fully contained
in $F_i$, it is homotopic to the core curve in the ring, and there
are two unique saturating curves $\C_\eta$ and $\C_{\eta + \delta}$
in $F_i$ such
that $\g\cap \C_\eta \not= 0$, and $\g \cap \C_{\eta + \delta} \not= 0$.

\noindent
$\underline{\hbox{Lemma 5.1}}.\,$ Consider a surface with an
admissible metric and a nontrivial
closed curve $\g$ that penetrates a ring domain $F_i$, isometric
to a flat cylinder of circumference $2\pi$, a distance $\delta$.
The length $l (\g )$ of the curve $\g$ satisfies the inequality
$$
l (\g ) \geq \, 2\pi \,\sqrt{1+ \delta^2/\pi^2}.\eqn\inequality
$$
The same inequality holds for a nontrivial closed curve $\g$
which extends a distance $\delta$ on $F_i$.

\noindent
$\underline{\hbox{Proof}}.\,$ Consider a curve $\g$ penetrating
$F_i$ a distance $\delta$. Then the two points $p_1$ and $p_2$ determine
a segment $\ov {p_1p_2}$ on $\C_0$ homotopic to the subcurve $\g'$ lying
completely on $F_i$ whose endpoints are $p_1$ and $p_2$. Let
$d$ denote the length on the (extension of the) flat metric on the
cylinder of the segment $\ov {p_1p_2}$. It is clear from the flat geometry
of the cylinder (see Fig.~3) that the length $l(\g')$ must exceed
that of the segment $\ov {p_1p_2}$ by
$$l(\g') - l(\ov {p_1p_2} ) \equiv e(d) \,
\geq \, 2 \, \sqrt{ \delta^2 + d^2/4} \, - d,\eqn\excess$$
Since the metric is admissible and the subcurve $\g'$ is homotopic
to the segment $\ov {p_1p_2}$, it follows that
$l(\g ) \geq 2\pi + e(d)$, where $e(d)$ is the excess length defined
above. The function $e(d)$ can be readily checked to be monotonically
decreasing in $d$. It will therefore attain its minimum value for
the maximum possible value of $d$, namely $2\pi$. We then
have
$$l(\g ) \geq 2\pi + e(2\pi ) = 2\pi \,\sqrt{1 + \delta^2 / \pi^2}.
\eqn\therewego$$
This is the desired bound. The computation for the case of a curve
that extends a distance $\delta$ in the flat cylinder $F_i$ is
enterely analogous, and the conclusion is the same. \hfill $\square$

\section{Proof of Amputation}

Consider a punctured
Riemann surface $\R$ equipped with a complete metric $\rho_0$
smooth near the punctures, and solving the
generalized minimal area problem. We have shown that around each
puncture $p_i$ there is a neighborhood where the metric is that
of a flat semiinfinite cylinder of circumference $2\pi$.

Let us now use this to find a canonical domain near $p_i$ on which
to perform the amputation.  Consider the set
$\U^*_i=\{p\in\R\mid$ there
exists exactly one nontrivial curve $\g_p$ through $p$ of length $2\pi$.
The curve $\gamma_p$ is homotopic to $p_i$,
and $\rho_0$ is smooth near $p\}$. Let $\U_i$ denote the connected
component of $\U^*_i$ which meets every neighborhood of $p_i$; the
domain $\U_i$ is foliated by geodesics homotopic to $p_i$, and is
conformally equivalent to the punctured disk $\{|\zeta_i|<1\}$.
This disk $\U_i$ with metric $\rho_0$ is isometric to a flat
semiinfinite cylinder of circumference $2\pi$ foliated by
saturating geodesics homotopic to $p_i$. It is the maximal
such cylinder around the puncture. The saturating
geodesics are the curves of constant $|\zeta |$. The disk $\U_i$
is the canonical domain defined by the minimal area metric.

{}From now on let us consider a single puncture $p$, and its corresponding
maximal disk $\U$ ($|\zeta | <1$).
Define on $\U$ the coordinate $t= -\log \zeta$,
and let $\C_\delta$ with
$\delta \in [0, \infty ]$ denote the
saturating geodesic $|t|=\delta$. The curve $\C_0$ is therefore the boundary
of $\U$, and $\delta$ measures the $\rho_0$-distance
between $\C_\delta$ and $\C_0$.
The curve $\C_\delta$ divides the surface $\R$ into two pieces:
the subdisk $\U_\delta$
($\equiv \U-\{ |\zeta| \geq e^{-\delta} \}$) and
the amputated surface $\R_\delta$ ($\equiv \R-\U_\delta$).
Finally we define, for a fixed $\delta \not= 0$ the ``stub''
$\stub_\delta$ to be the annulus $\U -\U_\delta$. Clearly
$\R_\delta = \R_0 \cup \stub_\delta$ (see Fig.~4).
\bigskip
\noindent
$\underline{\hbox{Theorem 5.1}}$. (Amputation) Consider a surface $\R$ with
a minimal area metric $\rho_0$ defining a canonical
domain around a puncture $p$. Assume the amputated surface $\R_\delta$
has a minimal area metric $\rho$ continuous in a neighborhood of
the boundary $\C_\delta$. Then $\rho$ is the restriction of $\rho_0$
to $\R_\delta$.
\medskip
\noindent
$\underline{\hbox{Proof}}$.
Our proof will involve several steps.
We will first assume there is a candidate metric on the
amputated surface having less area. The obvious idea, which
is to use the candidate metric on the amputated surface,
together with the original metric on the remainder of $\R$
to define an admissible metric on the whole surface, does
not work because the resulting metric would be discontinuous on the
cutting line and bad curves could appear. Nevertheless, with
a careful treatment of the neighborhood of the cutting line we will
succeed in constructing an admissible metric on $\R$ of lower area
that that of $\rho_0$.

We begin by assuming Theorem 5.1 does not hold and there is
another metric $\rho_1$
on $\R_\delta$ continuous on a neighborhood of $\C_\delta$,
satisfying all the length conditions on $\R_\delta$ and having
less area than $\rho_0$. Note that $\rho_1 \not= \alpha \rho_0$,
where $\alpha$ is a constant, since the lower area condition
would imply $\alpha <1$, and this would make the geodesics in the
stub $\stub_\delta$ shorter than $2\pi$.

Our aim will be to construct, using $\rho_0$ and $\rho_1$ a
candidate metric on $\R_\delta$, of area lower than $\rho_0$, which
can be glued back to the cylinder representing the puncture
giving an admissible metric on $\R$. This would be a contradiction
since it would show that $\rho_0$ was not the minimal area metric
on $\R$. Begin by considering the family $\rho_\epsilon$ of metrics
on $\R_\delta$ given by
$$
\rho_\epsilon = (1-\epsilon )\rho_0 + \epsilon \rho_1,\eqn\one
$$
where $0<\epsilon <1$. It is not difficult to see that
$\rho_0$ is an admissible metric on $\R_\delta$ (see the proof
of Corollary 5.7 below), and, by hypothesis, the metric $\rho_1$
is admissible, so, by linearity, we see that  $\rho_\epsilon$
is an admissible metric on $\R_\delta$. Let $A_0$ and $A_1$ denote
the area of $\R_\delta$ in the $\rho_0$ and $\rho_1$ metrics
respectively. We now calculate the area $A_\epsilon$ of $\R_\delta$
using the metric $\rho_\epsilon$. It follows from Eqn. \one\ that
$$
A_\epsilon = (1-\epsilon )^2 A_0
+ 2(1-\epsilon ) \epsilon \int_{\R_\delta} d^2\xi\, \rho_0 \rho_1
+ \epsilon^2 A_1.\eqn\two
$$
Since we know that $\rho_1 \not= \alpha \rho_0$ ($\alpha$ a constant),
the Schwarz inequality gives us
$$
\int_{\R_\delta} d^2\xi\, \rho_0 \rho_1 = \sqrt{A_0 A_1}
-\Delta ,\eqn\three
$$
with $\Delta >0$, since the Schwarz inequality is not saturated.
It follows from Eqns. \two\ and \three\ that
$$
A_\epsilon  = A_0 - |\beta_1 | \epsilon + |\beta_2 | \epsilon^2,
\eqn\four
$$
where the constants $|\beta_1 |$ and $|\beta_2 |$ are both strictly
positive and given by
$$
|\beta_1 | = 2 \left( A_0 (1- \sqrt{{A_1\over A_0}}) + \Delta \right)
\, >0,\quad |\beta_2 | = (\sqrt{A_0} -\sqrt{A_1})^2 + 2\Delta \,>0.
\eqn\five
$$
It follows from Eq. \four,  that for sufficiently small $\epsilon$ the
area $A_\epsilon$ is strictly smaller than $A_0$.
\medskip
We will work throughout with the canonical local coordinates
determined by the metric $\rho_0$; in these coordinates
$\rho_0 = 1$ on the cylinder.  Consider a neighborhood
$\N(r)\subset \R_\delta$ of the
boundary $\C_\delta$
consisting of all points in $\R_\delta$ whose $\rho_0$-distance
to $\C_\delta$ is smaller than $r$.
Choose $\epsilon$, such that $\epsilon^{1/4} < \delta$; therefore
the neighborhood $\N (\epsilon^{1/4} )$ is contained in the stub
$\stub_\delta$.
Let the constant $K$ be defined by
$$
1+ K \equiv \sup\limits_{ \N (\epsilon^{1/4})} \rho_1 ,\eqn\six
$$
where $\rho_1$ is the value of the conformal factor in the flat local
coordinates defined by $\rho_0$.
The constant $K$ is bounded since the candidate metric $\rho_1$
has been assumed to be continuous on a neighborhood of $\C_\delta$ and
therefore, for sufficiently small $\epsilon$ equation \six\ defines
a finite $K$. It follows from Eq.~\one\ that
$$
\sup\limits_{\N (\epsilon^{1/4})} \rho_\epsilon =
1+K\epsilon \equiv M(\epsilon ),\eqn\seven
$$
where we have defined $M(\epsilon )$ for later convenience.

Let us see that $K >0$. If $K < 0$ then, $\rho_1 <1$ throughout the
neigborhood, but then it would not be admissible. If $K = 0$, then
$\rho_1 \leq 1$ in the full neighborhood. Admissibility then requires
$\rho_1 \equiv 1$ and therefore $\rho_1 = \rho_0$ in the neighborhood.
The punctured disk could be restored at this stage, since the metric
is continuous along the cutting curve, which is of length $2\pi$, the
full metric is admissible, and of lower area. This is a contradiction.
Therefore $K >0$, and bounded, is the only case we need to consider.

Let us extend the definition of the metric $\rho_1$ to $\R$
by letting $\rho_1 = \rho_0$ on $\U_\delta$.
This metric is clearly discontinuous on $\R$
and may have bad curves. We quantify this possibility
in the following lemma.

\noindent
$\underline{\hbox{Lemma 5.2}}.\,$ For any nontrivial closed curve
$\gamma$ on $\R$, we have $l_{\rho_1} (\gamma) \geq 2\pi (1-K)$.
\medskip
\noindent
$\underline{\hbox{Proof}}.\,$  Any curve fully on $\R_\delta$ or
$\U_\delta$ clearly satisfies the inequality (in fact they are both
longer than $2\pi$). The only problem are the curves that cross
$\C_\delta$. Divide such a curve $\gamma$ into two pieces
$\gamma_{in}$, which is the part of $\gamma$ on $\R_\delta$, and
$\gamma_{out}$, which is the part of $\gamma$ on $\U_\delta$
(out of the amputated surface). The
possibility that curves can be shorter than $2\pi$ is due to the
discontinuity of the metric $\rho_1$ across $\C_\delta$. The portion
$\gamma_{out}$ is made of segments of lengths $b^i$, where the index
$i$ labels the different segments. Each segment is homotopic to a
segment $a_i$ on $\C_\delta$. The length of this segment
$a_i \in \C_\delta$ depends on
whether it is taken with respect to the metric $\rho_1$ in $\R_\delta$
(the in-metric) or with respect to the (extension of the
metric in $\U_\delta$ to the) metric in $\overline\U_\delta$ (the
out-metric). Let those lengths be
be denoted by $a^i_{in}$ or $a^i_{out}$ respectively.

Since the maximum value of $\rho_1$ in a neighborhood of
$\C_\delta$ lying on $R_\delta$ is $(1+K)$ (Eq.~\six ), it follows that
$$
a^i_{in} \leq
(1+K)a^i_{out} \quad \hbox{so that}
\quad a^i_{in} - a^i_{out} \leq Ka^i_{out}.
\eqn\eight
$$
We now have that
$$
l_{\rho_1}(\gamma) = l_{\rho_1}(\gamma_{in}) + \sum b^i
\geq l_{\rho_1}(\gamma_{in}) + \sum a^i_{out},\eqn\nine
$$
where the last inequality follows from $b^i \geq a^i_{out}$, which
follows because on $\U_\delta$ the metric $\rho_1$ is that of a flat
cylinder, and therefore a segment of a core curve ($a^i_{out}$)
is shorter than any other homotopic open curve between its
endpoints.

Let us now consider two cases. If $\sum a^i_{out} \geq 2\pi$, then
the above equation tells us we are done (the curve is
longer than $2\pi$). Let us now consider the case when
$\sum a^i_{out} \leq 2\pi$. It follows from Eq.~\nine\ that
$$
l_{\rho_1}(\gamma) \geq l_{\rho_1}(\gamma_{in}) +  \sum a^i_{in}
- \sum (a^i_{in} - a^i_{out} ) .\eqn\ten
$$
The first two terms in the right hand side give the length of
a closed curve entirely on $\R_\delta$ and of the same homotopy
type as the original curve. Since $\rho_1$ is admissible on $\R_\delta$
these two terms add up to
$2\pi$ or more. For the last term we use Eq.~\eight, and find
$$l_{\rho_1}(\gamma) \geq 2\pi -K\sum a^i_{out} .\eqn\elevenn$$
Since we are considering the case when $\sum a^i_{out} \leq 2\pi$,
the above inequality reduces to
$l_{\rho_1} (\gamma ) \geq 2\pi (1-K)$, which is the desired
statement.\hfill $\square$
\medskip
We now extend the definition of $\rho_\epsilon$
to the full surface $\R$ by letting
$\rho_\epsilon \equiv \rho_0$ on $\U_\delta$. This is
compatible with eqn.\one\ and the definition of $\rho_1$
over $\R$.
It is now possible to show that most curves are good
for the metric $\rho_\epsilon$ on $\R$, where we recall that a
$\rho_\e$-good curve has $\rho_\e$-length of at least $2\pi$.
This is the content
of the following Lemma:
\medskip
\noindent
$\underline{\hbox{Lemma 5.3}}.\,$ The metric $\rho_\epsilon$ in
$\R$ is admissible for all curves except for those completely contained
in $\N (\epsilon^{1/3}) \cup \U_\delta$, and crossing the boundary curve
$\C_\delta$.
\medskip
\noindent
$\underline{\hbox{Proof}}.\,$  Nontrivial closed curves are either
homotopic to the puncture or are not. Let us begin by showing that
the curves that are not homotopic to the puncture are always
good. If such curve is completely contained in $\R_\delta$ it
is clearly good (see below Eq.~\one ). Consider then a curve $\gamma$
that extends into $\U_\delta$ (see Fig.~4). Since it completely
penetrates the stub $\stub_\delta$ we have (Lemma 5.1)
$$
l_{\rho_0} (\gamma) \geq 2\pi \sqrt{1+
\delta^2/\pi^2},\eqn\twelve
$$
and moreover, $l_{\rho_1} (\gamma) \geq 2\pi
(1-K)$ (Lemma 5.2).  Therefore the equality
$$
l_{\rho_\epsilon} (\gamma) = (1-\epsilon) l_{\rho_0} (\gamma)
+ \epsilon l_{\rho_1} (\gamma ),\eqn\thirteen
$$
leads to the inequality
$$
l_{\rho_\epsilon} (\gamma) \geq
(1-\epsilon )\, 2\pi \sqrt{1+ \delta^2/\pi^2}
+ \epsilon \, 2\pi (1-K).\eqn\fourteen
$$
It follows that $l_{\rho_\epsilon}(\gamma ) \geq 2\pi$ if we have
$$
(1-\epsilon) (\sqrt{1+\delta^2/\pi^2} -1) -K\epsilon \geq 0.\eqn\fifteen
$$
Taking $\epsilon<1/2$, it is then sufficient to require
$$
\sqrt{1+\delta^2/\pi^2} -1 \geq 2K\epsilon
\quad \hbox{so that} \quad {\delta^2 \over 4 \pi^2}
\geq  K\epsilon + (K\epsilon )^2 .\eqn\sixteen
$$
This is easily satisfied.  If $(\delta^2/4\pi^2)< 1$, we take
$\epsilon < \delta^2/8K\pi^2$. If $(\delta^2/4\pi^2)\geq 1$, it
is sufficient to take $\epsilon < 1/2K$. This shows that curves
that are not homotopic to the puncture can be made to have
$\rho_\e$ lengths at least $2\pi$ by
choosing $\epsilon$ sufficiently small.

Now consider the curves homotopic to the puncture. If they lie
completely on $\R_\delta$ or $\U_\delta$ they are good. By the
statement of the lemma the only curves we need to discuss are those
that go into the surface $\R_\delta$ beyond $\N (\epsilon^{1/3})$ and
also get into $\U_\delta$. We can easily give an estimate of the
length of such curve in the $\rho_0$ metric. Since it penetrates a
foliation for a distance $\epsilon^{1/3}$ we have that
$$
l_{\rho_0} (\gamma ) \geq 2\pi \sqrt{1 + \epsilon^{2/3}/\pi^2} ,
\eqn\seventeen
$$
and using our estimate on $l_{\rho_1}(\gamma )$, we find
$$
l_{\rho_\epsilon} (\gamma) \geq
(1-\epsilon ) \, 2\pi \sqrt{1+ \epsilon^{2/3}/\pi^2}
+ \epsilon \, 2\pi (1-K).\eqn\eighteen
$$
Admissibility requires that (again $\epsilon<1/2$)
$$
\sqrt{1+ \epsilon^{2/3}/\pi^2} \geq 1+2K\epsilon \quad
\hbox{so that} \quad
{\epsilon^{2/3}\over 4\pi^2} \geq K\epsilon + (K\epsilon )^2.
\eqn\nineteen
$$
Since the left hand side of the last inequality is less than one, it
is sufficient to take $K\epsilon <\epsilon^{2/3}/8\pi^2$, which
just requires $\epsilon <1/(8\pi^2)^3K^3$. Thus for sufficiently
small $\epsilon$ we have admissibility. This concludes our proof
of lemma 5.3.\hfill $\square$
\medskip
We now need to improve the metric $\rho_\epsilon$ in order to get
the remaining curves to have sufficient length. These are the curves
that cross the boundary $\C_\delta$, and on $\R_\delta$ do not extend
beyond a distance $\epsilon^{1/3}$ of $\C_\delta$.
Define now on $\R_\delta$ a new metric ${\rho_\epsilon}'$ that is
flat and constant near the curve $\C_\delta$:
$$
{\rho_\epsilon}' = \rho_\epsilon + f_\epsilon (r)
(M(\epsilon ) - \rho_\epsilon ) , \eqn\twenty
$$
where $M(\e )$ was defined in \seven , and $r$ denotes $\rho_0$-distance
to $\C_\delta$.
Here the function $f_\epsilon$ is an interpolating function whose
value is $f_\epsilon (r) = 1$ for $r <\epsilon^{1/3}$, and
$f_\epsilon (r) = 0$ for $r> \epsilon^{1/4}$. For
$\epsilon^{1/3} <r< \epsilon^{1/4}$, the function $f_\epsilon$ is
monotonically decreasing and continuous. Note that on
$\N (\epsilon^{1/3} )$ we have ${\rho_\epsilon}' = M(\epsilon )
=1+K\epsilon$, which is just a constant. On $\U_\delta$ we set
the metric ${\rho_\epsilon}' \equiv \rho_0$.
\medskip
\noindent
$\underline{\hbox{Lemma 5.4}}.\,$ The metric ${\rho_\epsilon}'$ is an
admissible metric on $\R$, and its area,  for sufficiently
small $\epsilon$, is lower than that of $\rho_0$.
\medskip
\noindent
$\underline{\hbox{Proof}}.\,$ The metric ${\rho_\epsilon}'$
is admissible on $\R$ for the same curves $\rho_\epsilon$ was, since
it differs from $\rho_\epsilon$ by the addition of a term that
is always positive. The remaining curves,
namely those that extend over the two domains
$\N (\epsilon^{1/3})$ and $\U_\delta$ only, now clearly are
$\rho_\e'$-good.
Since over all of $\N (\epsilon^{1/3})$ the
metric ${\rho_\epsilon}' = 1+ K\epsilon > 1= \rho_0$, any closed
curve could only have grown in size with respect to its original total
length in the $\rho_0$ metric. Thus any such curve must be longer
or equal to $2\pi$. The improvement term in Eq.~\twenty\ was necessary
to get this type of curves to be $\rho_\e '$-good.

Let us now calculate the area
${A'}_\epsilon$ of the metric ${\rho_\epsilon}'$ on $R_\delta$.
It follows from Eq.~\twenty\ that:
$$
{A'}_\epsilon = A_\epsilon
+ 2\int d^2 \xi \rho_\epsilon f_\e (M(\epsilon ) - \rho_\epsilon )
+ \int d^2 \xi f_\epsilon^2 (M(\epsilon ) - \rho_\epsilon )^2.
\eqn\twentyone
$$
Since supp($f_\e$)$\subset \N (\e^{1/4})$, we see that
we need only estimate these integrals over
$\N (\epsilon^{1/4} )$. In this region we have the following
inequalities
$$
f_\epsilon (r) \leq 1,\quad \rho_\epsilon \leq 1+ K\epsilon,\quad
M(\epsilon ) - \rho_\epsilon \leq 1 + K\epsilon - (1-\epsilon) =
(K+1)\epsilon ,\eqn\twentytwo
$$
where in the last one, we have used the fact that the lowest
possible value of $\rho_1$ is zero. We can now estimate the integrals
in Eq.~\twentyone\ to find ($\int d^2\xi = 2\pi \int dr$)
$$
{A'}_\epsilon \leq A_\epsilon
+ 4\pi (1+K\epsilon ) (1+ K) \epsilon^{5/4}
+ 2\pi (1+K)^2 \epsilon^{9/4} .\eqn\twentythree
$$
Taking $K\epsilon <1$, namely $\epsilon <1/K$ and using Eq. \four, we
have
$$
{A'}_\epsilon \leq  A_0 - |\beta_1 | \epsilon
+|\eta| \epsilon^{5/4} + |\beta_2| \epsilon^2
+|\kappa| \epsilon^{9/4},\eqn\twentyfour
$$
where $|\eta | = 8\pi (1+ K)$ and $|\kappa |=2\pi (1+K)^2$. It is
clear from Eq.~ \twentyfour\ that for sufficiently small $\epsilon$ the
area of ${\rho_\epsilon}'$ on $\R_\delta$ is lower than that of
$\rho_0$ on $\R_\delta$. The same is therefore true for the area
on the whole surface $\R$. This concludes our proof of
lemma 5.4. \hfill$\square$

Having established Lemma 5.4, the existence of a metric $\rho_1$
of lesser area than $\rho_0$ on $\R_\delta$ has enabled us to
construct an admissible metric on $\R$ of area lower than that
of the minimal area metric $\rho_0$. This contradiction establishes
that the restriction of $\rho_0$ is indeed the minimal area metric
on the amputated surface $\R_\delta$.
This concludes our proof of Theorem 5.1. \hfill $\square$

\section{Plumbing Minimal Area Metrics}

We conclude this section by describing the minimal area metric on a
surface obtained by the conformal plumbing of surfaces which admit
minimal area metrics which are smooth and complete near the
appropriate punctures. What follows is a detailed
exposition of the ideas briefly sketched in [Zw2].

To begin, we recall from the introduction that the process of
conformal plumbing begins either with a surface $\R_0$ with at
least a pair of punctures, or two surfaces $\R'_0$ and $\R''_0$
each with at least one puncture.
In the first case, say, we consider small
neighborhoods $U_{|t|,1}=\{\zeta_1\mid |\zeta_1|<|t|\}$ and
$U_{|t|,2}=\{\zeta_2\mid |\zeta_2|<|t|\}$ of the punctures $p_1$
and $p_2$ (respectively) and then, for $|t|<\epsilon$, we form
the (complex) identification space
$\R_t=(\R_0 - (U_{t,1}\cup U_{t,2}))/\left<\zeta_1\zeta_2=t\right>$.
The space $\R_t$ is a Riemann surface of genus one more than the
genus of $\R_0$ but with two fewer punctures, and is said to be the
result of the conformal plumbing of $\R_0$. A similar operation
forms, from $\R'_0$ and  $\R''_0$, a new surface $\R'_t$ with genus
the sum of the genera of $\R'_0$ and $\R''_0$ and with two fewer
punctures than the total of those on $\R'_0$ and $\R''_0$.

We now describe the minimal area metric on $\R_t$ in terms of the
minimal area metric on $\R_0$; the case of the minimal area metric
on $\R'_t$ is analogous. So let $\rho_0$ be the minimal area metric
hypothesized throughout this paper; if $\rho_0$ is smooth and
complete near the punctures $p_1$ and $p_2$, then we have seen that
$\rho_0$ is flat in some neighborhood of $p_i$ and foliated there by
geodesics of length $2\pi$.

We perform our
plumbing using the canonical domain $\U_i$ around the puncture
$p_i$, defined by the minimal
area metric (see our discussion at the beginning of \S5.1).
The domain $\U_i$ is foliated by geodesics homotopic to $p_i$, and is
conformally equivalent to the punctured disk $\{|\zeta_i|<1\}$.
We take $\U_{|t|,i}$ to be the domain parametrized by the subdisk
$\U_{|t|,i}=\{|\zeta_i|<|t|\}$ (note a small change in notation
from \S5.1).

Next we form, as above, the identification space
$\R_t=\R_0 - (\U_{|t|,1}\cup\U_{|t|,2})/
\left<\zeta_1\zeta_2=t\right>$. Now, the surface $\R_t$ admits the
alternative description as the smooth surface obtained by gluing
together the ends of
$\R^{\cut}_{|t|} =
\R_0 - (\U_{|t|^{1/2},1}\cup\U_{|t|^{1/2},2})$ corresponding to
$p_1$ and $p_2$, the gluing taking place along the curves
$|\zeta_1|=|t|^{1/2}$ and $|\zeta_2|=|t|^{1/2}$ and the
identification being made so that $\zeta_1\zeta_2=t$ along the seam
$\Gamma=\{|\zeta_1|=|t|^{1/2}=|\zeta_2|\}$. We observe that the
curves $|\zeta_1|=|t|^{1/2}$ and $|\zeta_2|=|t|^{1/2}$ are
geodesics in the metric $\rho_0$ which is flat near the seam, so
that the restriction of the metric $\rho_0$  to
$\R_0-\left(\ov{\U_{|t|^{1/2},1}}\cup\ov{\U_{|t|^{1/2},2}}\right)$
extends to a smooth metric $\rho_{0,t}$ on $\R_t$ which is flat near
the seam.

\medskip
\noindent
$\underline{\hbox{Corollary 5.7}}.\,$ For $|t|<e^{-2\pi}$, the
plumbed surface $\R_t$ admits a minimal area metric $\rho_t$ and
$\rho_t=\rho_{0,t}$.
\medskip
\noindent
$\underline{\hbox{Proof}}.\,$ We first observe that $\rho_{0,t}$ is
a candidate metric on $\R_t$. This is clear once we consider the
homotopically non-trivial curves on $\R_t$. If $\gamma$ is such
a curve then either $\gamma$ meets the seam $\Gamma$ or it does not.
If $\gamma$ does not meet $\Gamma$, then we may take $\gamma$ to be
a curve on $\R_0$.  Moreover, this curve $\gamma \in \R_0$ must be
non-trivial, and hence have length at least $2\pi$: if $\gamma$
is trivial in $\R_0$ it must bound a disk $D\subset\R_0$;
but then $D$ is contained in  $\R^{\cut}_{|t|}$ since $D$ cannot contain the
punctures and $\g = \partial D$ does not meet the seam $\G$. In that case
actually $D \subset \R_t$ and $\g$ would be trivial in $\R_t$, in
contradiction with the initial assumption that $\g$ was nontrivial.
If $\gamma$ meets the seam
$\Gamma$ and one of the arcs $\{|\zeta_i|=1\}$, then $\gamma$ must
either cross the union of the domains
$\{ |\zeta_1| \geq |t|^{1/2} \} \cup \{ |\zeta_2| \geq |t|^{1/2}\}$
or one of those subdomains twice, in either case acquiring a
$\rho_{0,t}$ length of at least $2\pi$, as long as $|t|<e^{-2\pi}$.
If $\gamma$ meets the seam $\Gamma$ but is properly contained in a
neighborhood of the seam of size $\pi$, then $\gamma$ must be homotopic
to the seam, and one sees easily that
such a curve has a length of at least the length of $\Gamma$, or $2\pi$.

Next we see that $\rho_{0,t}$ has least area among admissible
metrices on $\R_t$. From Theorem~5.1 (Amputation), we see that
$\rho_{0,t}$ is the minimal area metric on
$\R^{\cut}_{|t|}$. Then,
if there were an admissible metric $\rho_t$ on $\R_t$ with lower area
than $\rho_0$, then the metric $\rho_t$ would restrict to be a
metric on $\R^{\cut}_{|t|}$ of area lower than $\rho_{0,t}$. Yet the
metric $\rho_t$ would still be admissible for the minimal area
problem on $\R^{\cut}_t$, since any homotopically non-trivial curve $\g$
on $\R^{\cut}_t$ is a nontrivial curve on $\R_t$. To see this, either
use Van Kampen's theorem, or more concretely,
suppose $\g$ is trivial in $\R_t$, then it bounds a disk $D \in R_t$;
since $\g$ does not meet $\G$ on $\R^{\cut}_t$ (this is easily arranged
by a small deformation) either $\G$ is fully contained in $D$ or it
is completely disjoint from $D$. However, $\G$ cannot be contained in $D$,
since $\G$ is a nontrivial curve, therefore $D$ is a proper disk in
$\R^{\cut}_t$, and this implies that $\g$ is trivial in $\R^{\cut}_t$
in contradiction with the assumption that it was a nontrivial curve.
Thus the metric $\rho_t$ would be both admissible and of area lower
than $\rho_{0,t}$ on $\R^{\cut}_t$. This contradiction shows
that $\rho_{0,t}$ is of minimal area on $\R_t$, concluding the
argument. \hfill $\square$
\medskip
\noindent
$\underline{\hbox{Remark}}.\,$
This corollary provides for the existence of solutions to
the minimal area problem for surfaces of higher genus using
minimal area metrics of lower genus through the process of
plumbing. The resulting Riemann surfaces correspond to a
deleted neighborhood of the compactification divisor of $\ov{\M}_{g,n}$.
For example, it was shown in [Zw1] that the
minimal area problem was solvable for all punctured spheres
$\R = S^2 - \{p_1, \dots, p_n\}$, the solution being
given as the norm of a holomorphic Jenkins-Strebel quadratic
differential on $\R$.
This corollary then provides for the existence of solutions to the
minimal area problem of surfaces obtained by the plumbing of a
number of punctured spheres. Of course, in this particular case
one sees that these solutions are also given by Jenkins-Strebel quadratic
differentials.

\chapter{Minimal Area Metrics not Arising from Quadratic Differentials}

In this section we will give an example of a minimal area metric
solving the generalized minimal area problem. The unusual aspect
of this minimal area metric is that it does not arise from a
quadratic differential. This will be manifest since we will get
negative curvature singularities corresponding to an excess angle
of $\pi /2$. In quadratic differentials the excess (or defect) angle
must be an integer multiple of $\pi$ (an $n$-th order zero corresponding
to an excess angle of $n\pi$). Moreover the pattern of foliations
by geodesics of lengths $2\pi$ is also novel. For a minimal area
metric arising from a Jenkins-Strebel quadratic differentials one can
always take the horizontal trajectories of the differential to define
the foliating geodesics. Then the surface is completely foliated by
geodesics that do not intersect (geodesics ovelap along the
critical graph of the quadratic differential, this graph, of course,
has zero measure). In the example to be discussed it is not possible
to cover the surface with saturating geodesics that do not intersect.
This is probably the crucial feature of all the metrics solving the
generalized minimal area problem which do not
arise from quadratic differentials.

Let us now describe the surface and its metric. The surface will
turn out to be a genus five surface $\R$ with no punctures. It will
be constructed by gluing together two identical tori $\T$ and $\T'$
each with four boundaries. Each torus with boundaries is given by a
$2\pi$ by $2\pi$ square region in the $z$ plane, with the natural
flat metric $\rho_0 = 1$ on it, and with opposite edges identified to give
a torus. We cut four square holes on each torus, each square of perimeter
$2\pi$ and symmetrically centered, as shown in Fig.~5(a).
These are the four boundaries. The tori $\T$ and $\T'$ are joined by
four short flat tubes $\C_i$ (not shown in the figure) attached to
the boundary components. Each tube is a cylinder of circumference
$2\pi$ and height $\pi/2$. Note that at the corners of the square
holes the metric $\rho_0$ has curvature singularities with an excess
angle of $\pi /2$.

Let us verify that the metric is admissible. All nontrivial curves
homotopic to curves lying completely in $\T$ or $\T'$ are good
because they are longer than or have lengths equal to some
homotopic curve lying completely
in $\T$ (or $\T'$) and such curves are manifestly good.
Curves homotopic to a core curve in one of
the short tubes also belong to this class.
We also need to consider simple closed Jordan curves
going from $\T$ to $\T'$. These curves must travel along the tubes
an even number of times. Since the tubes have height $\pi /2$
the only case we need to consider is that of curves going across
two times, thus acquiring at least length $\pi$.
Suppose they go up and down the same tube. If we delete the two
segments going up and down the tube we obtain two open curves. Each
curve is longer than or equal to an open curve of the same homotopy type
that does not enter the tube in question again. Consider those
open curves, which are defined on $\R-$(tube). At least one of them should
be a nontrivial open curve since otherwise the original closed  curve
is either trivial or homotopic to a core curve in the tube. A nontrivial
open curve, however, must be larger than $3\pi /2$ (check the figure).
Thus we exceed the necessary length. The only curves left to consider
are those that meet two different tubes. Since the distance between the
different boundary components in $\T$ and $\T'$ is $\pi /2$ the curve
must gain an extra $\pi$ of length, and therefore must be good.

Before showing the metric is of minimal area let us describe the
pattern of foliations. These are indicated in the figure and
are of three types. The first type is foliations lying completely
in $\T$ or in $\T'$. There are four bands of foliations and they
cross, covering $\T$ (or $\T'$) once in some regions and twice
in others (Fig.~5(b)). The second type of foliations are those extending
both in $\T$ and $\T'$. They are indicated in Fig.~5(c) and
the geodesics go from one boundary component in $\T$ to
another, then up the tube to $\T'$, then to another boundary
component in $\T'$ and then down the tube to $\T$. Such curves
have length $2\pi$. They extend over the regions of
$\T$ and $\T'$ where the first type of foliations gave a
single covering. They also cover the tubes once.
There are eight bands of this type.
The third type of foliations are those whose geodesics are homotopic to
the core curves in the tubes, there are four such bands, one in each
tube (not shown in the figure). The three types of foliations put
together give a double covering of the complete surface $\R$.

Let us now prove that the metric is of minimal area. Beurling's
criterion (see [Ah]) applied to our problem says that a
metric $\rho_0$
solves the generalized minimal area problem if it is
admissible and there is a family of nontrivial closed
curves $\Gamma_0$ such that $l_{\rho_0} (\g ) = 2\pi$ for
all $\g \in \G_0$, and for any real valued $h$ in $\R$
such that
$$\int_\gamma h |dz| \geq 0 ,\eqn\conditionh$$
for all $\gamma \in \Gamma_0$, we have that
$$\int\hskip-6pt\int_{\R} h \rho_0 dxdy \geq 0.\eqn\desired$$

Let us show our metric on $\R$ satisfies this criterion.
The family $\Gamma_0$ consists of the three types of foliations
discussed above. Each
foliation covers a annular region $\R_i$ of the surface,
isometric to a flat strip of length $2\pi$ with edges identified.
Put rectangular coordinates $x,y$ on each strip, and let the
geodesics correspond to the constant $x$ lines.
It then follows from \conditionh\ that
$$\int dy\,\int dx h \geq 0, \quad \hbox{and so} \quad
\int\hskip-6pt\int_{\R_i} h \rho_0 dxdy \geq 0,\eqn\eachfol$$
since $\rho_0 = 1$. Adding over all foliations
we obtain
$$\sum_i \int\hskip-6pt\int_{\R_i} h \rho_0 dxdy
= 2 \int\hskip-6pt\int_{\R} h \rho_0 dxdy \geq 0,\eqn\hereitis$$
since all foliations together give
a double covering of the surface. This concludes our proof
that the metric $\rho_0$ in $\R$ is of minimal area.

\noindent
$\underline{\hbox{Remark}}.\,$
There is an interesting one parameter deformation of the above
metric [Ro]. We can vary the heights of the short tubes
continuously but in doing so we must keep the heights of
diametrically opposite tubes equal, and the
sum of heights of neighboring tubes equal to $\pi$. All
these metrics are of minimal area.
The endpoint of this deformation is a configuration where two
of the tubes collapse and the other two tubes become of
height $\pi$ each. This metric does not arise from a quadratic
differential nor we can cut the tubes and still have a
minimal area metric. (Had the tubes been longer than $2\pi$ we
could have cut them and obtain a minimal area metric.)
\medskip
\ack
The authors benefitted from interesting conversations with
R. Forman, T. Maskawa, K. Ranganathan and M. Rocek.
\bigskip

\noindent{\bf REFERENCES}
\medskip
\item{[Ah]} Ahlfors, L. V.: Conformal Invariants, topics
in geometric function theory. New York:
McGraw Hill, 1973.
\item{[Ber]} Bers, L.: Spaces of Degenerating Riemann Surfaces.:
Annals of Math Studies, No.79, Princeton University Press,
Princeton, NJ, 1974, 43--55.
\item{[Beu]} Beurling, A.: Collected Works of A. Beurling. Eds.
L. Carleson et.al. Boston: Birkhauser, 1989.
\item{[EL]} Eells, J.,  Lemaire, L.: Deformations of Metrics
and Associated Harmonic Maps, Patodi Memorial Vol. Geometry and
Analysis (Tata Inst., 1981), 33--45.
\item{[EM]} Earle, C.J., Marden, A.: Geometric Complex Coordinates
for Teichmuller Space. To appear.
\item{[Ga]} Gardiner, F.: Teichmuller theory and quadratic
differentials. New York: John Wiley 1987.
\item{[Ha]} Harer, J.: The cohomology of the moduli space of
curves, in Theory of Moduli: Lectures given at C.I.M.E. Springer
Lecture Notes in Mathematics, 1337. Springer Verlag, Berlin, 1988.
\item{[H]} Hartman, P: Ordinary Differential Equations.
Wiley, Baltimore, 1973.
\item{[Hi]} Hicks, N.: Notes of Differential Geometry, Van Nostrand,
Princeton, 1965.
\item{[Je]} Jenkins, J. A.: On the existence of certain general
extremal metrics. Ann. Math. {\bf 66} 440 (1957).
\item{[KKS]} T. Kugo, H. Kunitomo and K. Suehiro, Nonpolynomial Closed
String Field Theory. Phys. Lett. {\bf 226B} 48 (1989)
\item{[KS]} T. Kugo and K. Suehiro.: Nonpolynomial closed string
field theory: action and gauge invariance. Nucl. Phys. {\bf B337} 434 (1990)
\item{[Mas]} Masur, H.: The Extension of the Weil-Petersson
metric to the boundary of Teichmuller Space.  Duke Math. {\bf J43}
(1979), 623--635.
\item{[Ra]} Ranganathan, K.: A criterion for flatness in minimal
area metrics that define string diagrams. Comm. Math. Phys., to appear.
\item{[Ro]} M. Rocek, private communication.
\item{[SaZw]} M. Saadi and B. Zwiebach.: Closed string field
theory from polyhedra.  Ann. Phys. {\bf 192} 213 (1989).
\item{[SoZw]} H. Sonoda and B. Zwiebach.: Closed string field theory
loops with symmetric factorizable quadratic differentials.
Nucl. Phys. {\bf B331} 592 (1990).
\item{[Sp]} Spivak, M.: A Comprehensive Introduction to
Differential Geometry.:  Publish or Perish Press, Berkeley, 1979.
\item{[St]} Strebel, K.: Quadratic Differentials. New York:
Springer-Verlag, 1984.
\item{[W]} Wolf, M.: Infinite Energy Harmonic Maps and
Degeneration of Hyperbolic Surfaces in Moduli Space.
J. Diff. Geom. {\bf 33}, 487 (1991),
\item{[Zw1]} Zwiebach, B.:  How covariant closed string
theory solves a minimal area problem.:
Comm. Math. Phys. {\bf 136}, 83 (1991);
Phys. Lett. {\bf B241}, 343 (1990).
\item{[Zw2]} Zwiebach, B.: Quantum closed strings from minimal
area.: Mod. Phys. Lett. {\bf A5}, 2753 (1990).
\item{[Zw3]} Zwiebach, B.: Minimal area problems for quantum
open strings. Comm. Math. Phys. {\bf 141}, 577 (1991).
\item{[Zw4]} Zwiebach, B.: Recursion relations in closed string
field theory, Proceedings of the ``Strings 90'' Superstring
Workshop. Eds. R. Arnowitt, .et al. (World Scientific, 1991)
pp. 266-275.
\item{[Zw5]} Zwiebach, B.: Quantum closed string field theory:
Action and BRST transformations.  In preparation.
\endpage
\baselineskip 12pt
\noindent
{\bf Figure Captions}
\medskip
\noindent
$\underline{\hbox{Figure 1}}\,.$ The segments $A_0$ and ${A'}_0$
make up the $\rho_0$-geodesic
$\g_0$. The segments $A_\e$ and ${A'}_\e$ make up the
$\wh\rho_\e$-geodesic. The segments $A_0$ and $A_\e$ are homotopic
rel$\{ p_1 ,p_2 \}$. Interpolating between $A_0$ and $A_\e$ we show
some $\wh\rho_t$-geodesics $A_t$, where $\wh\rho_t = \rho_0 + th$.
\medskip
\noindent
$\underline{\hbox{Figure 2}}\,.$ We show a collection of
geodesics $\g (\cdot , t )$ with $t \in [0,\kappa ]$.
The geodesics are seen to converge because the
variation field $V(s) = {\partial \over \partial t} \g (s,t )$ is
not constant: ${\partial \over \partial s} \| V(s) \| \not= 0$ (at $q$).
The domains $R_+$ and $R_-$ where the deformation of the metric
is supported, are indicated.
\medskip
\noindent
$\underline{\hbox{Figure 3}}\,.$  We show a curve $\g$ that
penetrates a ring domain $F_i$, isometric to a flat cylinder
of circumference $2\pi$, a distance $\delta$.
To the right we open up the flat annulus
bounded by $\C_0$ and $\C_\delta$.
\medskip
\noindent
$\underline{\hbox{Figure 4}}\,.$  A minimal area metric complete
and smooth near a puncture determines a canonical domain $\U$ around the
puncture. The boundary of $\U$ is $\C_0$. This canonical domain
is isometric to a flat semiinfinite cylinder. The curve $\C_\delta$,
a distance $\delta$ away from $\C_0$, divides the surface
$\R$ into $\R_\delta$ and $\U_\delta$.
Also indicated is the stub $S_\delta$. The
amputation theorem relates the minimal area metric on the
amputated surface $\R_\delta$ to the original minimal area
metric on $\R$.
\medskip
\noindent
$\underline{\hbox{Figure 5}}\,.$ A genus five surface $\R$ with
a minimal area metric that does not arise from a quadratic differential
is built by joining together two tori, each with four boundaries, using
four short tubes.
(a) A torus $\T$ (or $\T'$): the edges of the square domain are
identified and the four square holes are the boundaries.
(b) The pattern of foliations that are completely contained in $\T$
(or $\T'$). (c) A partial view of the foliations that extend both
in $\T$ and $\T'$.
\end